\documentclass[11pt]{article}
\usepackage{mathrsfs}
\usepackage[all,pdf]{xy}
\usepackage{amsfonts}
\usepackage{lscape}
\usepackage{amssymb}
\usepackage{bbm}
\usepackage{color}
\usepackage{amsfonts,amssymb, mathrsfs, amsmath,   amssymb,  theorem,  float}
\usepackage{graphicx}  

\newtheorem{theorem}{Theorem}[section]

\newtheorem{cor}[theorem]{Corollary}
\newtheorem{lemma}[theorem]{Lemma}
\newtheorem{alg}[theorem]{Algorithm}
\newtheorem{remark}[theorem]{Remark}

\def\qed{\hfil {\vrule height5pt width2pt depth2pt}}

\def\qed{\hfil {\vrule height5pt width2pt depth2pt}}
\def\bref#1{(\ref{#1})}

\def\qed{\hfil {\vrule height5pt width2pt depth2pt}}

\def\C{\mathcal{C}}

\def\proof{{\noindent\em Proof.\,\,}}

\def\res{\hbox{\rm{Res}}}

\def\GCD{\hbox{\rm{gcd}}}

\def\bref#1{(\ref{#1})}

\def\N{{\mathbb N}}
\def\Q{{\mathbb Q}}

\def\C{{\mathbb C}}

\def\bref#1{(\ref{#1})}

\def\Z{{\mathbb Z}}
\def\C{{\mathbb C}}

\def\+{ \oplus}
\def\-{\ominus}
\def\*{\otimes}

\def\deg{\hbox{\rm{deg}}}

\def\modd{\hbox{\rm{mod}}}

\begin{document}



\title{Sparse Rational Function Interpolation  with\\ Finitely Many Values for the Coefficients\thanks{Partially supported by a grant from NSFC No.11688101.}}
\author{Qiao-Long Huang and Xiao-Shan Gao \\
 KLMM, UCAS,  Academy of Mathematics and Systems Science\\
 Chinese Academy of Sciences, Beijing 100190, China}
\date{}

\maketitle

\begin{abstract}
\noindent
In this paper, we give new sparse  interpolation algorithms for
black box univariate and multivariate rational functions $h=f/g$
whose coefficients are integers with an upper bound.
The main idea is as follows: choose a proper integer $\beta$
and let $h(\beta) = a/b$ with $\gcd(a,b)=1$.
Then $f$ and $g$ can be computed by solving the polynomial interpolation
problems $f(\beta)=ka$ and $g(\beta)=ka$ for some integer $k$.
It is shown that the univariate interpolation algorithm is almost optimal and
multivariate interpolation algorithm has low complexity in $T$ but
the data size is exponential in $n$.
\end{abstract}

\section{Introduction}
The interpolation for a sparse multivariate rational function  $h=f/g$ given as a black box
is a basic computational problem \cite{112,9,dima-1,10,3}.
%
%
Here, sparse means that an upper bound for the number of terms in $f$ and $g$
is given.
In many interpolation algorithms, an upper bound for the degrees of $f$ and $g$ is also given.
In \cite{huang-p}, a new constraint in sparse interpolation is considered:
it is assumed that the coefficients of a sparse polynomial are taken from a known finite set.
In this case, the polynomial can be recovered from the evaluation at one large sample point.
In this paper, we extend polynomial sparse interpolation under this assumption to rational functions.

In this paper, we consider the interpolation of
$h=f/g$, where $f,g\in \Z[x_1,x_2,\dots,x_n]$, and  $T, D, C$ are upper bounds for the
terms, degrees, and the absolute values of the coefficients of $f$ and $g$, respectively.
The main idea of the algorithm is reduce the interpolation of $h$ into
that of $f$ and $g$.

In the univariate case, let $\beta$ be a positive integer,
$h(\beta) = a/b,\gcd(a,b)=1$, and $\mu=\gcd(f(\beta),g(\beta))$.
If we can find $\mu$, then $f$ and $g$ can be recovered from
$f(\beta)= \mu a$ and $g(\beta) = \mu b$ by polynomial interpolations.
%
%
We prove that if $\beta \ge 2TC^2+1$, then for $k\in\N$, $k=\mu$ if and only if
there exist $p,q\in\Z[x]$ such that
$p(\beta) = k a$, $q(\beta) = k b$, and the coefficients of $p$ and $q$ are bounded by $C$.
Thus we can find $\mu$ by computing univariate polynomials
$p(\beta) = k a$, $q(\beta) = k b$ for $k=1,2,\ldots$ and check whether the coefficients
of $p$ and $q$ are bounded by $C$.
The value for $\beta$ can be further reduced in two ways.
If we evaluate $h$ at two sample points $h(\beta)$ and  $h(\beta+1)$,
then $\beta$ can be taken as $\beta = \sqrt{2T}C+1$.
For $\beta = 2C+1$, we can obtain a probabilistic algorithm.

In the multivariate case, the similar idea is used to give a probabilistic algorithm.
The sample point used is
$\beta_1 = (\beta+c_1)$,
$\beta_2 = (\beta+c_2)^{2D+1}$,
$\ldots$,
$\beta_n = (\beta+c_n)^{(2D+1)^{n-1}}$,
where  $\beta = 2TC^2+1$, and $c_1< c_2 < \cdots < c_n$ are random numbers.
We show that with high probability, we can recover $h$ from $h(\beta_1,\ldots,\beta_n)$.
%
%

The arithmetic complexity of the univariate interpolation is $\mathcal{O}(\mu T \log^2D)$ and the length
of the data is $\mathcal{O}(D(\log C + \log T))$.
The arithmetic complexity of the multivariate interpolation is $\mathcal{O}(\mu nT \log^2D)$ and the length of the data is $\mathcal{O}(D^n\log (TC^2+N)$.

Extensive experiments are done for the algorithms. It is shown that the univariate interpolation
algorithm is almost optimal in the sense that the time for interpolating $f/g$
is almost the same as that of interpolating $f$ and $g$.
For the multivariate case, the algorithm is less sensitive for $T$ but is
quite sensitive for $D$ and $n$ due to the fact that the sample data is of height $D^n$.

The rest of this paper is organized as follows.
In Section 2, we give some preliminary results.
In Section 3, we give interpolation algorithms for univariate sparse rational functions.
In Section 4, we give interpolation algorithms for multivariate rational sparse functions.
In section 5, experimental results are presented.
Conclusions are presented in Section 6.

\section{Preliminary algorithms on univariate polynomial interpolation}
In this section, we will present  some preliminary algorithms 
which will be used in the rest of this paper.
We always assume
\begin{equation}\label{eq-f}
f(x)=c_1x^{d_1}+c_2x^{d_2}+\dots+c_tx^{d_t}\end{equation}
where $d_1,d_2,\dots,d_t\in \N,d_1<d_2<\cdots<d_t$, and $c_1,c_2,\cdots,c_t\in A$, where $A\subset \mathbb{C}$ is a finite set.
Introduce the following notations
\begin{equation}\label{eq-e}
C:=\max_{a\in A}(|a|),\quad \varepsilon:=\min (\varepsilon_1,\varepsilon_2)
\end{equation}
where
$\varepsilon_1 :=\min_{a,b\in A,a\neq b}|a-b|$ and $\varepsilon_2 :=\min_{a\in A,a\neq 0} |a|$.
With these notations, we have

\begin{theorem}[\cite{huang-p}]\label{the-5}
If $\beta\geq \frac{2C}{\varepsilon}+1$, then $f(x)$ can be uniquely determined by $f(\beta)$.
\end{theorem}
Algorithms based on the above theorem were given in \cite{huang-p}.
In particular, the following interpolation algorithm for polynomials in $\Q[x]$ is given in \cite{huang-p}, which is needed in this paper.
\begin{alg}[UPolySIRat]\cite[Algorithm 2.14]{huang-p}\label{alg-uprc}
\end{alg}
{\noindent\bf Input:} $H,C\in\N$, $\beta=2CH(H-1)$, and a black box polynomial $f(x)\in \Q[x]$ whose coefficients
are in
\begin{equation}\label{eq-rat}
A=\{\frac b a \,|\, 0<a\leq H,|\frac{b}{a}|\leq C,a,b\in \Z\}.
\end{equation}

{\noindent\bf Output:} The exact form of $f(x)$.

\begin{theorem}\label{the-11}\cite{huang-p}
The arithmetic complexity of Algorithm \ref{alg-uprc} is $\mathcal{O}(t\log H)$ and the bit complexity is $\mathcal{O}((t\log H)(t\log H+d\log C+d\log H))$, where $d=\deg(f)$  and $t=\#f$.
\end{theorem}

The following results will be needed in this paper.

\begin{lemma}\cite{huang-p}\label{the-1}
If $\beta\geq 2C+1$, then
$|\frac{f(\beta)}{\beta^k}|=\begin{cases}
>\frac{1}{2},&\text{if }  k\leq d_t\\
<\frac{1}{2},&\text{if }  k> d_t
\end{cases}
$
\end{lemma}

As a consequence, we can compute the degree of $f(x)$ as follows.
\begin{lemma}\cite{huang-p}\label{lm-1}
If $\beta\geq \frac{2C}{\varepsilon}+1$, then
$d_t=\lfloor\log_\beta 2|f(\beta)|\rfloor$.
\end{lemma}

We now show how to find the lowest degree $d_1$ of $f(x)$.
We denote $\mathbf{mod}(a,b)$ to be the value $a \mod b$, where $b$ is a positive integer and $\mathbf{mod}(a,b)$ is in $\{0,\dots,b-1\}$.
We first check if $\mathbf{mod}(f(\beta),\beta)\neq 0$. If it does, then the lowest degree of $f(x)$ is $0$. Otherwise, we compute a list $[\beta,\beta^2,\beta^{2^2},\cdots,\beta^{2^s}]$, such that  $\mathbf{mod}(f(\beta),\beta^{2^s})=0$ and $\mathbf{mod}(f(\beta),\beta^{2^{s+1}})\neq0$. As $\beta^{2^i}=\beta^{2^{i-1}}\cdot \beta^{2^{i-1}}$, we need $O(s)$ arithmetic operations to obtain the list. Denote $B_{up}:=2^{s+1}, B_{down}:=2^s$. Then we know $B_{down}\leq d_1<B_{up}$, if $B_{up}-B_{down}=1$, then $d_1=B_{down}$.

If $B_{up}-B_{down}\neq1$, let $a:=\frac{|f(\beta)|}{\beta^{B_{down}}}$ and check if $\mathbf{mod}(a,\beta)\neq 0$. If it does, then $d_1
=B_{down}$. Otherwise,  we also use the list to divide $a$ one by one until finding the integer $s_1$ which satisfies that $\mathbf{mod}(a,\beta^{2^{s_1}})=0$ and $\mathbf{mod}(a,\beta^{2^{s_1+1}})\neq0$.
Since $\frac{a}{\beta^{2^{s_1}}}=\frac{|f(\beta)|}{\beta^{2^s+2^{s_1}}}$ and $\mathbf{mod}(a,\beta^{2^{s_1}})=0$, $\beta^{2^s+2^{s_1}}< \beta^{2^{s+1}}$, this implies $s_1\leq s-1$.
Now we can update the upper and lower bound, $B_{up}:=B_{down}+2^{s_1+1}$ and $B_{down}:=B_{down}+2^{s_1}$, so $B_{up}-B_{down}=2^{s_1}$.
Now we check again if $B_{up}-B_{down}=1$ or $\mathbf{mod}(a,\beta)\neq0$. If it does, then $d_1=B_{down}$. Otherwise  update $a:=\frac{a}{\beta^{2^{s_1}}}$ and find $s_2$.

Repeating the above procedure to determine $s_2, s_3,s_4,\cdots$. Since $\log_2 d_1\geq s>s_1>s_2>s_3>\cdots$, the procedure will stop when some $s_i=0$. After at most $\log_2 d_1+1$ iterations, we can find the integer $d_1$. The procedure needs $\mathcal{O}(\log^2 d_1)$ arithmetic operations.

In order to be used in the next section, the input of  Algorithm \ref{alg-deg}, Algorithm \ref{alg-10} and \ref{alg-UPolySIMod} are modified as follows: $f(\beta)$ is denoted as as $\rho$
and a variable $x$ is introduced.
In rational function interpolation, $\rho$ is $\frac{f(\beta)}{\mu}$ for some integer $\mu$. When $\mu=1$, Algorithm \ref{alg-UPolySIMod} always return the correct $f$.

\begin{alg}[MinDeg]\label{alg-deg}\,
\end{alg}

\noindent{\bf Input:} $\rho,\beta\in\N$.

{\noindent\bf Output:} The degree of the minimum monomial in $f(x)$.
\begin{description}
\item[Step 1:] Set $a:=\rho$.

\item[Step 2:] $\mathbf{If}$ $\mathbf{mod}(a,\beta)\neq 0$ $\mathbf{then}$ $\mathbf{return}$ $0$.

\item[Step 3:] Find $E:=[\beta,\beta^2,\dots,\beta^{2^s}]$ such that $\mathbf{mod}(a, \beta^{2^s})=0$ and $\mathbf{mod}(a,\beta^{2^{s+1}})\neq0$.

\item[Step 4:] Let $B_{up}:=2^{s+1}$, $B_{down}:=2^s$.

\item[Step 4:]$\mathbf{while}$ $(B_{up}-B_{down})>1$ $\mathbf{do}$

$a:=\frac{a}{\beta^{B_{down}}}$;

$\mathbf{if}$ $\mathbf{mod}(a,\beta)\neq 0$ $\mathbf{then}$ $\mathbf{return}$ $B_{down}$; $\mathbf{end\ if}$;

Find $s_1$ such that $\mathbf{mod}(a,\beta^{2^{s_1}})=0$ and $\mathbf{mod}(a, \beta^{2^{s_1+1}})\neq 0$;

Update $B_{up}:=B_{down}+2^{s_1+1}; B_{down}:=B_{down}+2^{s_1}$;

$\mathbf{end\ do}$;

\item[Step 6:] $\mathbf{return}$ $B_{down}$.
\end{description}

\begin{alg}[MinCoef]\label{alg-10}
\end{alg}

{\noindent\bf Input:} $\rho\in\C$, $\beta,d\in\N$, a variable $x$, an upper bound $C$ of $\|f(x)\|_\infty$.

{\noindent\bf Output:} The coefficient of $x^d$ in $f(x)$ or $0$.

\begin{description}
\item[Step 1:] $v:=\frac{\rho}{\beta^d} \mod \beta$;

\item[Step 2:]

$\mathbf{if}$ $C<v<\beta-C$ $\mathbf{then}$  $\mathbf{return}$ $0$; $\mathbf{end\ if}$;

$\mathbf{if}$ $v\leq C$ $\mathbf{then}$  $\mathbf{return}$ $v $; $\mathbf{end\ if}$;

$\mathbf{if}$ $v\geq \beta-C$ $\mathbf{then}$  $\mathbf{return}$ $(v-\beta) $;  $\mathbf{end\ if}$;

\end{description}

\begin{alg}[UPolySIMod]\label{alg-UPolySIMod}
\end{alg}

{\noindent\bf Input:} $\rho\in\C$, $\beta\geq2C+1$, a variable $x$, an upper bound $C$ of the coefficients of  $f(x)$.

{\noindent\bf Output:} The exact form of $f(x)$ or failure if some coefficients of $f(x)$ is larger than $C$.

\begin{description}
\item[Step 1:] Let $u:=\rho$.

\item[Step 2:]
$g:=0,k:=0$;

$\mathbf{while}$ $u\neq0$ $\mathbf{do}$

 $d:=\mathbf{MinDeg}(\rho,\beta)$;

 $c:=\mathbf{MinCoef}(u,\beta,d,x,C)$;

$\mathbf{if}$ $c=0$ $\mathbf{then}$ $\mathbf{return}$ $\mathbf{failure}$; $\mathbf{end\ if}$;

$g:=g+c\cdot x^{d+k}$;

$u:=\frac{u-c\beta^d}{\beta^{d+1}}$;

$k:=k+d+1$;

$\mathbf{end\ do}$;

\item[Step 3:] $\mathbf{return}$ $g$

\end{description}

\begin{theorem}\label{the-14}
The arithmetic complexity of Algorithm \ref{alg-UPolySIMod} is $\mathcal{O}(t\log^2D)$ and the bit complexity is $\mathcal{O}(td\log^2d\log \beta)$, where $t=\#f$, $d=\deg(f)$,
and $C$ an upper bound of $\|f(x)\|_\infty$.
\end{theorem}
\proof
As shown in the paragraph before Algorithm \ref{alg-deg},
it takes $\mathcal{O}(\log^2 d)$ arithmetic operations in domain $\Z$ to obtain the minimum degree of $f(x)$. Since $f(x)$ has $t$ terms, the arithmetic complexity is $\mathcal{O}(t\log^2d)$.
Since $|f(x)|\leq C(\beta^{d_t}+\cdots+\beta+1)=C\frac{\beta^{d_t+1}-1}{\beta-1}\leq \frac{1}{2}\beta^{d_t+1}$, the bit complexity is $\mathcal{O}(t\log^2d\log (\beta)^{d})=\mathcal{O}(td\log^2d\log \beta)$.\qed

\section{Univariate rational function interpolation}
In this section, we give several sparse interpolation algorithms
for univariate rational functions.

\subsection{A basic interpolation algorithm}
In this section, we give a polynomial-time deterministic interpolation algorithm which is the starting point for more efficient algorithms.

We first introduce several notations used in this section.
Denote $\Z(x)$ to be the rational functions $\frac{f(x)}{g(x)}$
such that $f(x),g(x)\in\Z[x]$ and $\gcd(f,g)=1$.
In this paper, for $f(x),g(x)\in \Z[x]$, $\gcd(f,g)$ also contains the
greatest common factor of the coefficients of $f$ and $g$.

Let $h(x)=\frac{f(x)}{g(x)}\in\Z(x)$ and the coefficients of $f$ and $g$ are in $$A=\{-C,-C+1,\cdots,1,1,\cdots,C\}.$$
Denote $\deg(h):=\max\{\deg(f),\deg(g)\},$
$\#h:=\max\{\#f,\#g\}$,
$\|h\|_\infty:=\max\{\|f\|_\infty,\|g\|_\infty\}$,
where $\#f$ is the number of the terms of $f$
and $\|f\|_\infty$ is the maximal absolute value of the coefficients of $f$.

For a positive integer $\beta$, let $h(\beta)=\frac{a}{b}$, where $a$ and $b$ are integers such that $\GCD(a,b)=1$.
Let $\mu=\gcd(f(\beta),g(\beta))>0$.  Then, we have
\begin{equation}\label{eq-mu}
a=\frac{f(\beta)}{\mu},b=\frac{g(\beta)}{\mu}
\end{equation}
Denote $f_1(x):=\frac{1}{\mu} f(x),g_1(x):=\frac{1}{\mu} g(x)$. Then  $f_1(\beta)=a,g_1(\beta)=b$, and the coefficients of $f_1(x),g_1(x)$ are in $\frac{1}{\mu} A=\{-\frac{C}{\mu},-\frac{C-1}{\mu},\cdots,-\frac{1}{\mu},\frac{1}{\mu},\cdots,\frac{C-1}{\mu},\frac{C}{\mu}\}$. If we can give an upper bound $H$ for $\mu$ and let $\beta\geq 2CH(H-1)+1$, then we can recover $f_1$ and
$g_1$ using the Algorithm \ref{alg-uprc} and hence $f/g=f_1/g_1$. Therefore, a key issue in sparse interpolation for rational functions is to determine an upper bound for $\mu$.
The following lemmas give such an estimation.

\begin{lemma}\cite[p.147]{8}\label{lm-r11}
Let $f,g\in \Z[x]$, and $n=\deg(f)\geq m=\deg(g) \geq 1$. Then

$|\res(f,g,x)|\leq \parallel f\parallel_2^m \parallel g\parallel_2^n\leq(n+1)^{m/2}(m+1)^{n/2}\parallel f\parallel _\infty ^m \parallel g\parallel _\infty ^n$, where $\res(f,g,x)$ is the Sylvester resultant of $f$ and $g$ wrt $x$.
\end{lemma}

\begin{lemma}\label{lm-bd1}
If $f,g\in \Z[x]$, and $D\geq \max\{\deg(f),\deg(g)\}$, $C\geq \max\{\|f\|_\infty,\|g\|_\infty\}$, then we have $\mu \leq (D+1)^D C^{2D}$, where $\mu$ is defined in \bref{eq-mu}.
\end{lemma}
\proof
Since $\gcd(f(x),g(x))=1$, $\res(f,g,x)\ne0$.
By \cite[p.147]{8}, there exist two nonzero polynomials $s(x),t(x)\in \Z[x]$, such that $f(x)s(x)+g(x)t(x)=\res(f,g,x)$. So we have $f(\beta)s(\beta)+g(\beta)t(\beta)=\res(f,g,x)$.
Since $\res(f,g,x)$ is an integer, we have $\gcd(f(\beta),g(\beta))|\res(f,g,x)$.
By Lemma \ref{lm-r11},   $\mu \le |\res(f,g,x)|\leq (D+1)^D C^{2D}$.
\qed

\begin{theorem}
Let $h(x)=\frac{f(x)}{g(x)}$ with $D\geq \deg(h)$ and $C\geq \|h\|_\infty$. Denote $H:=(D+1)^D C^{2D}$. If $\beta\geq 2CH(H-1)+1$, then we can recover $h(x)$ from $h(\beta)$.
\end{theorem}

\proof
Use the notations in \bref{eq-mu}.
If we can interpolate the polynomials $\frac{1}{\mu}f(x),\frac{1}{\mu}g(x)$ from the values $a$ and $b$, then we finish the interpolation. By Lemma \ref{lm-bd1}, we know $|\mu|\leq|\res(f(x),g(x))|\leq H$.
Since the coefficients of $\frac{1}{\mu}f,\frac{1}{\mu}g$ are chosen in the finite set $A:=\{\frac{s}{t}|0<t\leq H,-C\leq  s\leq C,s,t\in \Z\}$,  $\varepsilon=\frac{1}{H(H-1)}$, when $\beta\geq 2CH(H-1)+1$, we can interpolate $\frac{1}{\mu}f(x),\frac{1}{\mu}g(x)$ from $a,b$ with Algorithm \ref{alg-uprc}.
Thus, $h(x)$ can be recovered from $h(\beta)$.\qed

We now give the algorithm.
\begin{alg}[URFunSI0]\label{alg-urf1}
\end{alg}

{\noindent\bf Input:} A black box  $h\in \Z(x)$, $D,C\in\N$, where $D\geq \deg(h),C\geq \|h\|_\infty$.

{\noindent\bf Output:} The exact form of $h(x)$.

\begin{description}
\item[Step 1:] Let $H:=(D+1)^D C^{2D}, \beta:=2CH(H-1)+1$.

\item[Step 2:] Evaluate $h(\beta)$, assume $h(\beta)=\frac{a}{b}$.

\item[Step 3:] Let
$f(x):=\mathbf{UPolySIRat}(a,\beta,C,H)$ and
$g(x):=\mathbf{UPolySIRat}(b,\beta,C,H)$.

\item[Step 5:] Return $\frac{f(x)}{g(x)}$.
\end{description}

\begin{theorem}\label{the-12}
The arithmetic operations of Algorithm \ref{alg-urf1} are $\mathcal{O}(TD\log C+TD\log D)$ and the bit complexity is $\mathcal{O}((TD\log C+TD\log D)(TD\log C+TD\log D+D\log C+D^2\log C+D^2\log D)$.
\end{theorem}
\proof
By Theorem \ref{the-11}, the arithmetic complexity of Algorithm $\mathbf{UPolySIRat}$ is
$\mathcal{O}(T\log H)$ and bit complexity is $\mathcal{O}((T\log H)(T\log H+D\log C+D\log H))$. Since $H=(D+1)^D C^{2D}$ and we call Algorithm $\mathbf{UPolySIRat}$ twice,
the theorem follows immediately.\qed

It should be pointed out that Theorem \ref{the-12} is a theoretical result,
since the number $\beta$ is too large. Practical algorithms will be given
in the following sections, which are modifications of variants of
Algorithm \ref{alg-urf1}.

\subsection{Deterministic incremental interpolation}
In Algorithm \ref{alg-urf1}, we use an upper bound for $\mu$.
In this section, an algorithm will be given, where $\mu$ will be searched incrementally.
We first give a lemma.

\begin{lemma}\label{the-8}
Assume $f_1,f_2,g_1,g_2\in\Z[x]$, and $\gcd(f_1,g_1)=1,\deg(f_2)\leq \deg(f_1),\deg(g_2)\leq \deg(g_1)$. If $\frac{f_1}{g_1}=\frac{f_2}{g_2}$, then there exists a nonzero integer $\delta$, such that $f_2=\delta f_1,g_2=\delta g_1$.
\end{lemma}
\proof
Since $\frac{f_1}{g_1}=\frac{f_2}{g_2}$, we have $f_1g_2=g_1f_2$ and hence $f_1|g_1f_2$. Since $\gcd(f_1,g_1)=1$,  we have $f_1|f_2$. From $\deg(f_2)\leq \deg(f_1)$, there exists a rational number $\frac{a}{b}$, such that $f_2=\frac{a}{b}f_1$. For the same reason we have $g_2=\frac{a}{b}g_1$.
Since $f_2,g_2\in \Z[x]$,  all their coefficients are integers. So $b$ divides all the coefficients of $f_1,g_1$, as $\gcd(f_1,g_1)=1$, and hence  $b=\pm 1$. So $\delta=\frac{a}{b}$ is an integer.\qed

\begin{theorem}\label{the-6}
Assume that $h(x)=\frac{f(x)}{g(x)}\in \Z(x)$ with  $T\geq \#h,C\geq \|h\|_\infty$. If $\beta\geq 2TC^2+1$, then we can recover $h(x)$ from $h(\beta)$.
\end{theorem}
\proof
%
Let $a,b,\mu$ be introduced in \bref{eq-mu}.
We claim that for $i=1,2,\cdots$, only when $i=\mu$, the values $a\cdot i,b\cdot i$ correspond to two polynomials with coefficients bounded by $C$.
We prove the claim by contradiction.
Assume there exists an $i_0<\mu$, such that $a\cdot i_0, b\cdot i_0$ corresponding to two polynomials $f_1(x),g_1(x)$ in $\Z[x]$ with $C\geq \|f_1\|_\infty,\|g_1\|_\infty$.
Since $|i_0a|<|\mu a|,|i_0b|<|\mu b|$, we have $\deg(f_1)\leq \deg(f),\deg(g_1)\leq \deg(g)$ by Lemma \ref{lm-1}.
Then we have $\frac{f(\beta)}{g(\beta)}=\frac{f_1(\beta)}{g_1(\beta)}$.
This can be changed into $f(\beta)g_1(\beta)=f_1(\beta)g(\beta)$. If we let $w(x):=f(x)g_1(x)$, $v(x):=f_1(x)g(x)$, then $w(\beta)=v(\beta)$. Since $T\geq \#f,\#g$,  $TC^2\geq\|w\|_\infty,\|v\|_\infty$.
Since $\beta\geq 2TC^2+1$, we have $w(x)=v(x)$, which can be changed into $\frac{f(x)}{g(x)}=\frac{f_1(x)}{g_1(x)}$. By the Lemma \ref{the-8}, we have $f_1(x)=\delta f(x),g_1(x)=\delta g(x)$, where $\delta$ is a nonzero integer, then $|i_0a|=|f_1(\beta)|=|\delta f(\beta)|\geq |f(\beta)|=|\mu a|$. This is a contradiction, so we prove the theorem.\qed

Theorem \ref{the-6} leads to the following deterministic algorithm.
\begin{alg}[URFunSI1]\label{alg-urf2}
\end{alg}

{\noindent\bf Input:} A black box  rational function $h(x)\in \Z(x)$, $T,C$, where $T\geq\#h,C\geq\|h\|_\infty$.

{\noindent\bf Output:} The exact form of $h(x)$.

\begin{description}
\item[Step 1:] Let $\beta:=2TC^2+1$.

\item[Step 2:] Evaluate $h(\beta)$, assume $h(\beta)=\frac{a}{b}$.

\item[Step 3:] $i=1$;

\item[Step 4:]

$f:=\mathbf{UPolySIMod}(a\cdot i,\beta,x,C)$;

$\mathbf{if}$ ($f=failure$ $\mathbf{or}$ $\#f>T$)  $\mathbf{then}$ $i:=i+1$; go to Step $4$; $\mathbf{end\ if}$

\item[Step 5:]

$g:=\mathbf{UPolySIMod}(b\cdot i,\beta,x,C)$;

$\mathbf{if}$ ($g=failure$ $\mathbf{or}$ $\#g>T$) $\mathbf{then}$ $i:=i+1$; go to Step $4$; $\mathbf{end\ if}$

\item[Step 6:]
$\mathbf{return}$ $\frac{f}{g}$.
\end{description}


\begin{theorem}\label{the-22}
The arithmetic complexity of Algorithm \ref{alg-urf2} is $\mathcal{O}(\mu T\log^2 D)$, and the height of the data is $\mathcal{O}(D(\log C+\log T))$,
where $\mu$ is defined in \bref{eq-mu}. In particular, when $\mu=1$, the arithmetic complexity is $\mathcal{O}(T\log^2 D)$.
\end{theorem}
\proof
The analysis of arithmetic complexity is similar to that of Theorem \ref{the-14}.
Since $\beta=2TC^2+1$, $f(\beta)$ is $\mathcal{O}(C(TC^2)^D)$ and the height of the data is $\mathcal{O}(D(\log C+\log T))$.\qed

\subsection{Deterministic incremental interpolation with two points}
In Algorithm \ref{alg-urf2}, we recover $h(x)$ from $h(\beta)$ for $\beta=2TC^2+1$.
In this section, we show that $h(x)$ can be recovered from $h(\beta)$ and $h(\beta+1)$
for a much smaller $\beta=\lceil\sqrt{2T}C\rceil$.
The following lemma shows how to recover a polynomial from two smaller points.

\begin{lemma}\label{the-7}
Let $f(x)=c_tx^{d_t}+c_{t-1}x^{d_{t-1}}+\cdots+c_1x^{d_1}\in\Z[x],d_1<d_2<\cdots <d_t$, and $C\geq \|f\|_\infty$.
If $\beta\geq \sqrt{2C}$, then $f(x)$ can be recovered from $f(\beta)$ and $f(\beta+1)$.
\end{lemma}
\proof
Assume that there exists another $g(x)=a_sx^{k_s}+a_{s-1}x^{k_{s-1}}+\cdots+a_1x^{k_1},k_1<k_2<\cdots <k_s$, and $C\geq \|g\|_\infty$, such that
$g(\beta)=f(\beta),g(\beta+1)=f(\beta+1)$.
Firstly, we prove $d_1=k_1$.
It is clear that $d_1$ ($k_1$) is the largest integer such that
${\modd}(f(\beta), \beta^{d_1})={\modd}(f(\beta+1),(\beta+1)^{d_1})=0$ (${\modd}(g(\beta), \beta^{k_1})=0,{\modd}(g(\beta+1),(\beta+1)^{k_1})=0$).
%
Since $f(\beta)=g(\beta),f(\beta+1)=g(\beta+1)$, we have $d_1=k_1$.

Next, we prove $a_1=c_1$.
From $\frac{f(\beta)}{\beta^{d_1}} \mod \beta=c_1$,
$\frac{g(\beta)}{\beta^{d_1}} \mod \beta=a_1$,
$\frac{f(\beta+1)}{(\beta+1)^{d_1}} \mod (\beta+1)=c_1$,
$\frac{g(\beta+1)}{(\beta+1)^{d_1}} \mod (\beta+1)=a_1$,
we have
\begin{equation*}
\begin{cases}
(a_1-c_1) \mod \beta=0\\
(a_1-c_1) \mod (\beta+1)=0
\end{cases}
\end{equation*}
Since $\gcd(\beta,\beta+1)=1$, we have $(a_1-c_1) \mod \beta(\beta+1)=0$.  $|a_1|,|c_1|\leq C$, so $|a_1-c_1|\leq 2C$. But $|\beta(\beta+1)|\geq \sqrt{2C}(\sqrt{2C}+1)> 2C$, so $a_1=c_1$.
The other terms can be proved by induction.\qed

\begin{theorem}
Assume $h(x)=\frac{f(x)}{g(x)}\in\Z(x)$, $T\geq \#h,C \geq \|h\|_\infty$. If $\beta\geq \lceil\sqrt{2T}C\rceil$, $h(x)$ can be recovered from $h(\beta)$ and $h(\beta+1)$.
\end{theorem}
\proof
Use the same notations as Theorem \ref{the-6}.
We still prove it by contradiction.
Assume there exists an $i_0<\mu$, such that $a\cdot i_0, b\cdot i_0$ correspond  to two integer polynomials with $C\geq \|f_1\|_\infty,\|g_1\|_\infty$.
Since $|i_0a|<|\mu a|,|i_0b|<|\mu b|$, we have $\deg(f_1)\leq \deg(f),\deg(g_1)\leq \deg(g)$.
Then we have $\frac{f(\beta)}{g(\beta)}=\frac{f_1(\beta)}{g_1(\beta)},\frac{f(\beta+1)}{g(\beta+1)}=\frac{f_1(\beta+1)}{g_1(\beta+1)}$. This can be change to $f(\beta)g_1(\beta)=f_1(\beta)g(\beta),f(\beta+1)g_1(\beta+1)=f_1(\beta+1)g(\beta+1)$.
Let $w(x):=f(x)g_1(x)$, $v(x):=f_1(x)g(x)$. Then $w(\beta)=v(\beta),w(\beta+1)=v(\beta+1)$.
Since $T\geq \max\{\#f,\#g\}$,   $TC^2\geq \max\{\|w\|_\infty,\|v\|_\infty\}$.
From $\beta\geq \lceil\sqrt{2T}C\rceil$, by Lemma \ref{the-7}, we have $w(x)=v(x)$, or $\frac{f(x)}{g(x)}=\frac{f_1(x)}{g_1(x)}$. By Lemma \ref{the-8}, the same reason as Theorem \ref{the-6}, we prove the theorem.\qed

Based on the above theorem, an interpolation algorithm using two points can be given.
In the following algorithm, we assume $T\geq5$. In this case, $\sqrt{2T}C\geq2C+1$, so the evaluation satisfies the input condition of Algorithm $\mathbf{UPolySIMod}$.
\begin{alg}[URFunSI2]\label{alg-urf3}
\end{alg}

{\noindent\bf Input:} A black box   $h(x)\in \Z(x)$, $T,C$, where $T\geq \# h,C\geq \|h\|_\infty$.

{\noindent\bf Output:} The exact form of $h(x)$.

\begin{description}
\item[Step 1:] Let $T_1:=\max(T,5)$, $\beta:=\lceil\sqrt{2T_1}C\rceil$.

\item[Step 2:] Evaluate $h(\beta),h(\beta+1)$ and assume $h(\beta)=\frac{a_1}{b_1},h(\beta+1)=\frac{a_2}{b_2}$.

\item[Step 3:] $i=1$;

\item[Step 4:]

$f:=\mathbf{UPolySIMod}(a_1\cdot i,\beta,C)$;

$\mathbf{if}$ ($f=failure$ $\mathbf{or}$ $\#f>T$) $\mathbf{then}$ $i:=i+1$; go to Step $4$; $\mathbf{end\ if}$

\item[Step 5:]

$g:=\mathbf{UPolySIMod}(b_1\cdot i,\beta,C)$;

$\mathbf{if}$ ($g=failure$ $\mathbf{or}$ $\#g>T$) $\mathbf{then}$ $i:=i+1$; go to Step $4$; $\mathbf{end\ if}$

\item[Step 6:]
$\mathbf{if}$ $\frac{f(\beta+1)}{g(p+1)}=\frac{a_2}{b_2}$ $\mathbf{then}$ return $\frac{f}{g}$; $\mathbf{else}$ $i:=i+1$; go to Step $4$.

\end{description}

\begin{theorem}
The arithmetic complexity of Algorithm \ref{alg-urf3} is $\mathcal{O}(\mu T\log^2 D)$, and the length of the data is $\mathcal{O}(D(\log C+\log T))$. In particular, when $\mu=1$, the arithmetic complexity is $\mathcal{O}(T\log^2 D)$.
\end{theorem}
\proof
The analysis of arithmetic complexity is the same as Theorem \ref{the-22}.\qed

Note that the complexity of Algorithm \ref{alg-urf3} is the same as that of
Algorithm \ref{alg-urf2}, but Algorithm \ref{alg-urf3} is practically much faster than
Algorithm \ref{alg-urf2} as shown in Section \ref{sec-exp}.

\subsection{Probabilistic univariate rational function interpolation}
In Algorithms \ref{alg-urf2} and \ref{alg-urf3},  $\beta=2TC^2+1$
and $\beta=\lceil\sqrt{2T}C\rceil$.
In this section, we will give a probabilistic algorithm
where  $\beta=3C+1$ under the condition that a degree bound for $f$ is known.

\begin{lemma}\label{lm-101}
Assume $h(x)=\frac{f(x)}{g(x)}\in\Z(x)$, $C\geq \|h\|_\infty, D\geq \deg(f)$.
Let $\beta\geq2C+1$, $h(\beta)=\frac{a}{b}$, and $\mu=\gcd(f(\beta),g(\beta))$.
Then $|\mu|\leq \lfloor\frac{\beta^{D+1}}{2|a|}\rfloor$.
\end{lemma}
\proof
By Lemma \ref{the-1}, $|\frac{f(\beta)}{\beta^{D+1}}|<\frac{1}{2}$.
Since $h(\beta)=\frac{a}{b}$, we have $a=\frac{1}{\mu}f(\beta)$, and $\frac{|a|}{\beta^{D+1}}=|\frac{\frac{1}{\mu} f(\beta)}{\beta^{D+1}}|<\frac{1}{2|\mu|}$.
Then we can give  an upper bound $|\mu|<\frac{\beta^{D+1}}{2|a|}$.
Since $\mu$ is an integer, $|\mu|\leq\lfloor\frac{\beta^{D+1}}{2|a|}\rfloor$.\qed

We can give a lower bound of degree of $f(x)$.
Assume $h(\beta)=\frac{a}{b}$.
By Lemma \ref{the-1}, the number $d$ satisfying $\frac{|a|}{\beta^d}>\frac{1}{2},\frac{|a|}{\beta^{d+1}}<\frac12$ is  a lower degree bound of $f(x)$.
The lower and upper degree bounds will avoid lots of computing.

In this subsection, we use two points $h(\beta),h(\beta+1)$ to interpolate $h(x)$. The following theorems will show some relations between the two points.

\begin{lemma}\label{lm-2}
Assume $f(x)=c_1x^{d_1}+c_2x^{d_2}+\cdots+c_tx^{d_t}\in \Z[x],C\geq \|f\|_\infty,d_1<d_2<\cdots<d_t$. Let $\beta\geq 2C+1$ and $Q:=\frac{f(\beta)/\beta^{d_t}}{f(\beta+1)/(\beta+1)^{d_t}}$, $E:=1+\frac{2C}{\beta(\beta-1)}$. Then we have
$\frac{1}{E}<Q< E$.
\end{lemma}
\proof
Denote $q_1:=\frac{c_1\beta^{d_1}+c_2\beta^{d_2}+\cdots+c_{t-1}\beta^{d_{t-1}}}{\beta^{d_t}}$
 and $q_2:=\frac{c_1(\beta+1)^{d_1}+c_2(\beta+1)^{d_2}+\cdots+c_{t-1}(\beta+1)^{d_{t-1}}}{(\beta+1)^{d_t}}$.
Then $Q=\frac{f(\beta)/\beta^{d_t}}{f(\beta+1)/(\beta+1)^{d_t}}=\frac{c_t+q_1}{c_t+q_2}=1+\frac{q_1-q_2}{c_t+q_2}$.
Since $|c_t|\geq 1,|q_2|<\frac{\varepsilon}{2}=\frac{1}{2}$, we have $|c_t+q_2|> \frac{1}{2}$. So $|\frac{q_1-q_2}{c_t+q_2}|<2|q_1-q_2|$. From
 \begin{align}
 |q_1-q_2|={} &|c_1 (\frac{1}{\beta^{d_t-d_1}}-\frac{1}{(\beta+1)^{d_t-d_1}})+\cdots+c_{t-1}(\frac{1}{\beta^{d_t-d_{t-1}}}-\frac{1}{(\beta+1)^{d_t-d_{t-1}}})|\notag\\
 \leq{}&C[(\frac{1}{\beta^{d_t}}-\frac{1}{(\beta+1)^{d_t}})+(\frac{1}{\beta^{d_t-1}}-\frac{1}{(\beta+1)^{d_t-1}})+\cdots+(\frac{1}{\beta}-\frac{1}{\beta+1})]\notag\\
 ={} &C[(\frac{1}{\beta^{d_t}}+\frac{1}{\beta^{d_t-1}}+\cdots+\frac{1}{\beta})-(\frac{1}{(\beta+1)^{d_t}}+\frac{1}{(\beta+1)^{d_t-1}}+\cdots+\frac{1}{\beta+1})]\notag\\
 ={} &C \frac{\frac{1}{\beta}-\frac{1}{\beta^{d_t+1}}}{1-\frac{1}{\beta}}-C\frac{\frac{1}{\beta+1}-\frac{1}{(\beta+1)^{d_t+1}}}{1-\frac{1}{\beta+1}}\notag\\
={} &C \frac{1-\frac{1}{\beta^{d_t-1}}+\frac{\beta-1}{(\beta+1)^{d_t}}}{\beta(\beta-1)}<\frac{C}{\beta(\beta-1)}\notag
\end{align}
We deduce $|Q-1|\leq 2|q_1-q_2|<\frac{2C}{\beta(\beta-1)}$, so we prove the first inequality.

Note that $\frac{1}{Q}=\frac{f(\beta+1)/(\beta+1)^{d_t}}{f(p)/p^{d_t}}=\frac{c_t+q_2}{c_t+q_1}=1+\frac{q_2-q_1}{c_t+q_1}$. We also have $|\frac{q_2-q_1}{c_t+q_1}|<\frac{2C}{\beta(\beta-1)}$. Then $|\frac{1}{Q}-1|<\frac{2C}{\beta(\beta-1)}$  and
\begin{equation*}
\begin{cases}
1-\frac{2C}{\beta(\beta-1)}<Q< 1+\frac{2C}{\beta(\beta-1)}\\
1-\frac{2C}{\beta(\beta-1)}<\frac{1}{Q}< 1+\frac{2C}{\beta(\beta-1)}
\end{cases}
\end{equation*}
So
\begin{equation*}
\begin{cases}
1-\frac{2C}{\beta(\beta-1)}<Q< 1+\frac{2C}{\beta(\beta-1)}\\
\frac{1}{1+\frac{2C}{\beta(\beta-1)}}<Q< \frac{1}{1-\frac{2C}{\beta(\beta-1)}}
\end{cases}
\end{equation*}
Since $1+\frac{2C}{\beta(\beta-1)}\leq \frac{1}{1-\frac{2C}{\beta(\beta-1)}}$ and $1-\frac{2C}{\beta(\beta-1)}\leq \frac{1}{1+\frac{2C}{\beta(\beta-1)}}$, we prove the lemma.
\qed

\begin{lemma}\label{lm-3}
Suppose $h(x)=\frac{f(x)}{g(x)}\in \Z(x)$ and $h(\beta)=\frac{a_1}{b_1},h(\beta+1)=\frac{a_2}{b_2}$, where $\gcd(a_1,b_1)=1,\gcd(a_2,b_2)=1$, $D\geq \deg(f)\geq d$. Assume $a_1=\frac{f(\beta)}{\mu_1},a_2=\frac{f(p+1)}{\mu_2}$ and denote $Q_1:=\frac{a_1/p^{d}}{a_2/(p+1)^d},Q_2:=\frac{a_1/p^{D}}{a_2/(p+1)^D}$,
and $E:=1+\frac{2C}{\beta(\beta-1)}$. Then
$|Q_1|\frac{1}{E}< |\frac{\mu_2}{\mu_1}|< |Q_2|E$.
\end{lemma}
\proof
Let $Q:=\frac{f(\beta)/\beta^{d_t}}{f(\beta+1)/(\beta+1)^{d_t}}$. Then we have
$Q_1=\frac{\mu_2}{\mu_1}\frac{f(\beta)}{f(\beta+1)}\frac{(\beta+1)^{d_t}}{\beta^{d_t}}\frac{\beta^{d_t-d}}{(\beta+1)^{d_t-d}}=Q\frac{\mu_2}{\mu_1}\frac{\beta^{d_t-d}}{(\beta+1)^{d_t-d}}$
and
$Q_2=\frac{\mu_2}{\mu_1}\frac{f(\beta)}{f(\beta+1)}\frac{(\beta+1)^{d_t}}{\beta^{d_t}}\frac{(\beta+1)^{D-d_t}}{\beta^{D-d_t}}=Q\frac{\mu_2}{\mu_1}\frac{(\beta+1)^{D-d_t}}{\beta^{D-d_t}}$.
By Lemma \ref{lm-2},
$\frac{1}{1+\frac{2C}{\beta(\beta-1)}}< Q< 1+\frac{2C}{\beta(\beta-1)}.$
Then
$|Q_1|< |\frac{\mu_2}{\mu_1}|\frac{\beta^{d_t-d}}{(\beta+1)^{d_t-d}}(1+\frac{2C}{p(p-1)})\Rightarrow |\frac{\mu_2}{\mu_1}|> |Q_1| \frac{(p+1)^{d_t-d}}{p^{d_t-d}}\frac{1}{1+\frac{2C}{p(p-1)}}\geq |Q_1| \frac{1}{1+\frac{2C}{p(p-1)}}$
and
$|Q_2|> |\frac{\mu_2}{\mu_1}|\frac{(\beta+1)^{D-d_t}}{\beta^{D-d_t}}\frac{1}{1+\frac{2C}{\beta(\beta-1)}}\Rightarrow |\frac{\mu_2}{\mu_1}|< |Q_2| \frac{\beta^{D-d_t}}{(p+1)^{D-d_t}}(1+\frac{2C}{p(p-1)})\leq |Q_2| (1+\frac{2C}{p(p-1)})$.\qed

It is easy to see that we have the best result if  $D=d=\deg(f(x))$.
\begin{cor}\label{lm-100}
If $|Q_1|\geq E$, then $|\mu_2|> |\mu_1|$. If $|Q_2|\leq\frac{1}{E}$, then $|\mu_2|<|\mu_1|$.
\end{cor}
\proof
By Lemma \ref{lm-3}, we have $|Q_1|\frac{1}{E}<|\frac{\mu_2}{\mu_1}|< |Q_2| E$, and the lemma follows from this.\qed

Now we give the algorithm.
\begin{alg}[URFunSIP]\label{ProbUniRFunInt}
\end{alg}

{\noindent\bf Input:} A black box $h(x)=\frac{f(x)}{g(x)}\in\Z(x)$, $D,C$, where $D\geq \deg(h),C\geq \|h\|_\infty$.

{\noindent\bf Output:}  The exact form of $h(x)$ or a wrong rational function.

\begin{description}

\item[Step 1:]
Let $\beta:=3C+1$.

\item[Step 2:]
Evaluate $h(\beta),h(\beta+1)$, and assume $h(\beta)=\frac{a_1}{b_1},h(\beta+1)=\frac{a_2}{b_2}$.

\item[Step 3:]
%
Let $d:=\max(\lfloor \log_\beta (2a_1)\rfloor, \lfloor \log_{\beta+1} (2a_2)\rfloor$ (due to Lemma \ref{lm-1}).

\item[Step 4:]
Let $k_1:=\lfloor\frac{\beta^{D+1}}{|a_1|}\rfloor,k_2:=\lfloor\frac{(\beta+1)^{D+1}}{|a_2|}\rfloor$.

$Q_1:=|\frac{a_1/\beta^d}{a_2/(\beta+1)^d}|$,
$Q_2:=|\frac{a_1/\beta^D}{a_2/(\beta+1)^D}|$,
$E:=1+\frac{2C}{\beta(\beta-1)}$.

\item[Step 5:]Let $i:=1$.

$\mathbf{If}$ $Q_1\geq E$ $\mathbf{then}$ goto step 6;
$\mathbf{If}$ $Q_2\leq \frac{1}{E}$ $\mathbf{then}$ goto step 7.
%
%
%

$\mathbf{If}$ $k_1<k_2$, $\mathbf{then}$ goto step 6;
$\mathbf{Else}$
goto step 7.


\item[Step 6:]

$\mathbf{while}$ $i\leq k_1$ $\mathbf{do}$

$\mathbf{if}$ the interval $(\frac{Q_1}{E}i, Q_2Ei )$ includes an integer,
 $\mathbf{then}$ $f:=\mathbf{UPolySIMod}(a_1\cdot i,\beta,C)$;

 $\mathbf{if}$ $f=failure$
 $\mathbf{then}$ $i:=i+1$; goto step 6; $\mathbf{end\ if}$;

 $g:=\mathbf{UPolySIMod}(b_1\cdot i,\beta,C)$;

  $\mathbf{if}$
  $g=failure$
   $\mathbf{then}$ $i:=i+1$; goto step 6; $\mathbf{end\ if}$;

 $\mathbf{if}$ $h(\beta+1)=\frac{f(\beta+1)}{g(\beta+1)}$, $\mathbf{then}$ return $\frac{f(x)}{g(x)}$.

\item[Step 7:]
$\mathbf{while}$ $i\leq k_2$ $\mathbf{do}$

$\mathbf{if}$ the interval $(\frac{1}{Q_2E}i,\frac{E}{Q_1}i)$ includes an integer

 $\mathbf{then}$ $f:=\mathbf{UPolySIMod}(a_2\cdot i,\beta+1,C)$

 $\mathbf{if}$
 $f=failure$
  $\mathbf{then}$ $i:=i+1$; goto step 7; $\mathbf{end\ if}$;

 $g:=\mathbf{UPolySIMod}(b_2\cdot i,\beta+1,C)$;

 $\mathbf{if}$
 $g=failure$  $\mathbf{then}$ $i:=i+1$; goto step 7; $\mathbf{end\ if}$;

 $\mathbf{if}$ $h(\beta)=\frac{f(\beta)}{g(\beta)}$  $\mathbf{then}$ return $\frac{f(x)}{g(x)}$.

\end{description}

\begin{theorem}
The algorithm is correct. The arithmetic complexity of
the algorithm is $\mathcal{O}(\mu D\log^2 D)$, where $\mu\leq (1+\frac{1}{3C})^{D-d+2}\min\{\mu_1,\mu_2\}$. The height of the data is $\mathcal{O}(D\log C)$. In particular, when $\mu=1$, the arithmetic complexity is $\mathcal{O}(T\log^2 D)$.
\end{theorem}
\proof
%
For convenience, we assume
$h(\beta) = a_1/b_1, \gcd(a_1,b_1)=1, f(\beta)=\mu_1 a_1, g(\beta)=\mu_1 b_1$,
$h(\beta+1) = a_2/b_2, \gcd(a_2,b_2)=1,  f(\beta+1)=\mu_2 a_2, g(\beta+1)=\mu_2 b_2$, and $\mu_1,\mu_2>0$.
The main idea of the algorithm is to find one of $\mu_1,\mu_2$, and thus the exact value $f(\beta)$ or $f(\beta+1)$. Since $\beta\geq 2C+1$, we can recover $f(x)$ and $g(x)$ by Algorithm \ref{alg-UPolySIMod}.
In the algorithm, we  use an incremental approach to find the probably smaller one in $\{\mu_1, \mu_2\}$. We give some simple criterions to compare which one is small due to
Corollary {\ref{lm-100}}.
We explain each step of the  algorithm below.

In step 1, we use $\beta=3C+1$ instead of $2C+1$. This  trick is used to avoid certain computing. For example, if $0<i<\mu_1$, then $ia_1<\mu_1 a_1=f(\beta)$. So when we apply Algorithm $\mathbf{UPolySIMod}(i a_1,\beta,C)$, it may return $failure$, since with high probability, one of the coefficients is not in $[-C,C]$. On the other hand, this will never happen when $\beta=3C+1$.


In step 3, we find a lower degree bound $d$ of $f(x)$.
In step 4, $k_1,k_2$ are the upper bounds of $\mu_1,\mu_2$ by Lemma \ref{lm-101}. $Q_1,Q_2, E$ are the quantities defined in Lemma \ref{lm-3}.

In step 5, if $Q_1\geq E$, by Lemma \ref{lm-3}, $\frac{\mu_2}{\mu_1}>\frac{|Q_1|}{E}\geq 1$, or $\mu_2>\mu_1$.
If $Q_2\leq \frac{1}{E}$, by Lemma \ref{lm-3}, $\frac{\mu_2}{\mu_1}<Q_2E\leq 1$, or $\mu_2<\mu_1$.
If both of them are not satisfied, then we just compare the bounds $k_1,k_2$ of $\mu_1,\mu_2$, respectively.

In step 6, we handle the case $\mu_2>\mu_1$.
We first use $h(\beta)$ to recover $h(x)$. We need to know the number $\mu_1$.
We let $i$ increases from $1$ to $k_1$ and $\mu_1$ is one of them.
We check three cases: (1) From Lemma \ref{lm-3}, we know $Q_1\frac{1}{E}< \frac{\mu_2}{\mu_1}< Q_2E$, so $Q_1\frac{1}{E}\mu_1< \mu_2< Q_2E\mu_1$. If the interval $(\frac{Q_1}{E}i, Q_2Ei )$ includes an integer, it could be $\mu_1$; if it does not, then $i$ cannot be $\mu_1$.
(2) If $f=failure$ or $g=failure$ (this is the reason why we choose $\beta\geq3C+1$), then we increase $i$ by one.
(3) If $h(\beta+1)=\frac{f(\beta+1)}{g(\beta+1)}$, then we return the result.
Note that the probabilistic property of the algorithm comes from here:
even if $h(\beta+1)=\frac{f(\beta+1)}{g(\beta+1)}$, we are not sure whether we have the correct $h$.
In step 7, we handle the case $\mu_1>\mu_2$, which is similar to step 6.

We now prove the bound of $\mu$. If $Q_1\geq E$ or $Q_2\leq \frac{1}{E}$, then it is easy to see that $\mu=\min\{\mu_1,\mu_2\}$.
So now we assume $Q_1<E$ and $Q_2> \frac{1}{E}$. Firstly, we have $E=1+\frac{2C}{\beta(\beta-1)}<1+\frac{1}{\beta}$
and $Q_1=Q_2 \frac{\beta^{D-d}}{(\beta+1)^{D-d}}$.
Since $Q_1\frac{1}{E}<\frac{\mu_2}{\mu_1}< Q_2 E$, $Q_2>\frac{1}{E}$ and $Q_1<E$, we have $\frac{\mu_2}{\mu_1}>Q_1\frac{1}{E}=Q_2 \frac{\beta^{D-d}}{(\beta+1)^{D-d}}\frac{1}{E}> \frac{\beta^{D-d}}{(\beta+1)^{D-d}}(\frac{1}{E})^2>\frac{\beta^{D-d+2}}{(\beta+1)^{D-d+2}}$. So $\mu_1<\frac{\beta^{D-d+2}}{(\beta+1)^{D-d+2}}\mu_2<(1+\frac{1}{3C})^{D-d+2}\mu_2$.
For the similar reason, we have $\mu_2<(1+\frac{1}{3C})^{D-d+2}\mu_1$. So we have $\mu\leq (1+\frac{1}{3C})^{D-d+2}\min\{\mu_1,\mu_2\}$.

The analysis of arithmetic complexity is similar to that of Theorem \ref{the-14}. Since the missing factor $\mu$ may destroy the sparse structure, we use $D$ instead of $T$.
Since $\beta=3C+1$, $f(\beta)$ is $\mathcal{O}(C^D)$ and the height of the data is $\mathcal{O}(D\log C)$.\qed
%

Since the upper bound for the degree is given, we can avoid lots of computing.

\section{Multivariate rational function interpolation}


\subsection{Multivariate polynomial interpolation with Kronecker substitution}
In this section, we will give an algorithm based on a variant Kronecker substitution
as the starting point for multivariate rational function interpolation algorithms.

In the rest of section, we assume that the variables are ordered as $x_1\prec x_2\prec\dots\prec x_n$, and the lexicographic monomial order will be used. Let $m=x_1^{k_1}x_2^{k_2}\cdots x_n^{k_n}$ be a monomial and $\beta_1,\beta_2,\cdots,\beta_n\in \N$. Then we denote $\widehat{m}:=\beta_1^{k_1}\beta_2^{k_2}\cdots \beta_n^{k_n}$.

\begin{lemma}\label{lm-5}
Suppose $m_1=x_1^{k_1}x_2^{k_2}\cdots x_n^{k_n}>m_2=x_1^{s_1}x_2^{s_2}\cdots x_n^{s_n}$ in lexicographic order and   $k_i,s_j\leq D,i,j=1,2,\cdots,n$. If $\beta_1>1$,  $\beta_2\geq \beta_1^{D+1},\cdots,\beta_i\geq \beta_{i-1}^{D+1},\cdots,\beta_n\geq \beta_{n-1}^{D+1}$, then $\widehat{m}_1>\widehat{m}_2$ and $\frac{\widehat{m}_2}{\widehat{m}_1}\leq \frac{1}{\beta_1}$.
\end{lemma}
\proof
As $m_1>m_2$, without loss of generality, assume $k_n>s_n$. Then we have
\begin{equation*}
\beta_1^{s_1}\beta_2^{s_2}\cdots \beta_n^{s_n}\leq \beta_1^D\beta_2^D\cdots \beta_{n-1}^D \beta_n^{k_n-1}, \beta_n^{k_n}\leq  \beta_1^{k_1}\beta_2^{k_2}\cdots \beta_n^{k_n}
\end{equation*}
It is sufficient to prove $\beta_1^D\beta_2^D\cdots \beta_{n-1}^D \beta_n^{k_n-1}<\beta_n^{k_n}$. Dividing $\beta_n^{k_n-1}$ on both sides, it is sufficient to prove $\beta_1^D\beta_2^D\cdots \beta_{n-1}^D<\beta_n$.
Since
$\beta_1\cdot \beta_1^D\beta_2^D\cdots \beta_{n-1}^D=\beta_1^{D+1}\beta_2^D\cdots \beta_{n-1}^D
\leq \beta_2^{D+1}\beta_3^D\cdots \beta_{n-1}^D\leq \cdots \leq \beta_{n-1}^{D+1}\leq \beta_n$,
we have
$\beta_1\cdot \beta_1^D\beta_2^D\cdots \beta_{n-1}^D\leq \beta_n$.
Since $\beta_1>1$,  $\beta_1^D\beta_2^D\cdots \beta_{n-1}^D< \beta_n$. So we have $\widehat{m}_1>\widehat{m}_2$.
Since $\frac{\widehat{m}_2}{\widehat{m}_1}=\frac{\beta_1^{s_1}\beta_2^{s_2}\cdots \beta_n^{s_n}}{\beta_1^{k_1}\beta_2^{k_2}\cdots \beta_n^{k_n}}$, we assume there exists an $i$ such that $k_{i+1}=s_{i+1},k_{i+2}=s_{i+2},\dots,k_n=s_n$,  and $k_i>s_i$. Then
%
$\frac{\widehat{m}_2}{\widehat{m}_1}=\frac{\beta_1^{s_1}\beta_2^{s_2}\cdots \beta_i^{s_i}}{\beta_1^{k_1}\beta_2^{k_2}\cdots \beta_i^{k_i}}\leq \frac{\beta_1^D\beta_2^D\cdots \beta_{i-1}^D \beta_i^{k_i-1}}{\beta_1^{k_1}\beta_2^{k_2}\cdots \beta_{i-1}^{k_{i-1}}\beta_i^{k_i}}=\frac{\beta_1^D\beta_2^D\cdots \beta_{i-1}^D}{\beta_i}\cdot \frac{1}{\beta_1^{k_1}\beta_2^{k_2}\cdots \beta_{i-1}^{k_{i-1}}}
\leq\frac{1}{\beta_1}$
\qed

\begin{lemma}\label{the-9}
Let $f=c_1m_1+c_2m_2+\cdots+c_tm_t\in \C[x_1,x_2,\dots,x_n]$, $m_1 < m_2 < \cdots < m_t$ in lexicographic order, all the coefficients are in the finite set $A$, $m_i=x_1^{d_{i,1}}x_2^{d_{i,2}}\cdots x_n^{d_{i,n}}$, $D\geq \deg(f)$.
Let $\beta_1\geq \frac{2C}{\varepsilon}+1,\beta_2\geq \beta_1^{D+1},\cdots,\beta_i\geq \beta_{i-1}^{D+1},\cdots,\beta_n\geq \beta_{n-1}^{D+1}$, where $C,\varepsilon$ are given in $(\ref{eq-e})$. Then for any $j=1,2,\dots,n$,

\begin{equation}
\frac{|f(\beta_1,\beta_2,\dots,\beta_n)|}{\beta_j^k\beta_{j+1}^{d_{t,j+1}}\beta_{j+2}^{d_{t,j+2}}\cdots \beta_n^{d_{t,n}}}=\begin{cases}
> \frac{\varepsilon}{2}, &\text{if } k\leq d_{t,j}\\
< \frac{\varepsilon}{2}, &\text{if } k> d_{t,j}

\end{cases}
\end{equation}

\end{lemma}
\proof
It is sufficient to show that if $k= d_{t,j}$, then $|\frac{f(\beta_1,\beta_2,\dots,\beta_n)}{\beta_j^k\beta_{j+1}^{d_{t,j+1}}\beta_{j+2}^{d_{t,j+2}}\cdots \beta_n^{d_{t,n}}}|> \frac{\varepsilon}{2}$;
if $k= d_{t,j}+1$, then $|\frac{f(\beta_1,\beta_2,\dots,\beta_n)}{\beta_j^k\beta_{j+1}^{d_{t,j+1}}\beta_{j+2}^{d_{t,j+2}}\cdots \beta_n^{d_{t,n}}}|<\frac{\varepsilon}{2}$.

First note
 $\frac{\widehat{m}_i}{\widehat{m}_t}=\frac{\widehat{m}_i}{\widehat{m}_{i+1}}\frac{\widehat{m}_{i+1}}{\widehat{m}_{i+2}}\cdots \frac{\widehat{m}_{t-1}}{\widehat{m}_t}\leq \frac{1}{\beta_1^{t-i}}$.
When $k=d_{t,j}$, we have
 \begin{align}
 |f(\beta_1,\beta_2,\cdots,\beta_n)|\geq{} & |c_t|\widehat{m}_t-C (\widehat{m}_{t-1}+\widehat{m}_{t-2}+\cdots+\widehat{m}_1)\notag\\
 ={} &\widehat{m}_t (|c_t|-C (\frac{\widehat{m}_{t-1}}{\widehat{m}_t}+\frac{\widehat{m}_{t-2}}{\widehat{m}_t}\cdots+\frac{\widehat{m}_1}{\widehat{m}_t}))\notag\\
\geq{} & \widehat{m}_t (|c_t|-C (\frac{1}{\beta_1}+\frac{1}{\beta^2_1}+\cdots+\frac{1}{\beta_1^{t-1}}))\notag\\
\geq{} & \widehat{m}_t(\varepsilon-\frac{C}{\beta_1-1}+\frac{C}{\beta_1^t-\beta_1^{t-1}})\notag\\
>{} &\frac{\varepsilon}{2}\widehat{m}_t\notag
\end{align}

So $\frac{|f(\beta_1,\beta_2,\dots,\beta_n)|}{\beta_j^k\beta_{j+1}^{d_{t,j+1}}\beta_{j+2}^{d_{t,j+2}}\cdots \beta_n^{e_{t,n}}}> \frac{\varepsilon}{2} \frac{\widehat{m}_t}{\beta_j^k\beta_{j+1}^{d_{t,j+1}}\beta_{j+2}^{d_{t,j+2}}\cdots \beta_n^{d_{t,n}}}\geq\frac{\varepsilon}{2}$.

When $k=d_{t,j}+1$,
\begin{align}
|f(\beta_1,\beta_2,\cdots,\beta_n)|\leq{} & C (\widehat{m}_t+\widehat{m}_{t-1}+\cdots+\widehat{m}_1)\notag\\
={} &C\widehat{m}_t(1+\frac{\widehat{m}_{t-1}}{\widehat{m}_t}+\cdots+\frac{\widehat{m}_1}{\widehat{m}_t})\notag\\
\leq{} & C \widehat{m}_t(1+\frac{1}{\beta_1}+\frac{1}{\beta^2_1}+\dots+\frac{1}{\beta_1^{t-1}})\notag\\
={} &C \widehat{m}_t\frac{\beta_1-\frac{1}{\beta_1^{t-1}}}{\beta_1-1}\notag\\
={} &\widehat{m}_t \frac{C}{\beta_1-1}(\beta_1-\frac{1}{\beta_1^{t-1}})\notag\\
\leq{} & \frac{\varepsilon}{2} \widehat{m}_t\beta_1-\frac{\varepsilon}{2}\widehat{m}_t\frac{1}{\beta_1^{t-1}}\notag
\end{align}
Clearly, $\frac{\widehat{m}_t\beta_1}{\beta_j^k\beta_{j+1}^{d_{t,j+1}}\beta_{j+2}^{d_{t,j+2}}\cdots \beta_n^{d_{t,n}}}\leq1$, so
$\frac{|f(\beta_1,\beta_2,\dots,\beta_n)|}{\beta_j^k\beta_{j+1}^{d_{t,j+1}}\beta_{j+2}^{d_{t,j+2}}\cdots \beta_n^{d_{t,n}}}< \frac{\varepsilon}{2}$.
\qed

\begin{lemma}\label{lm-202}
Let $f=f_1x_n^{d_1}+f_2x_n^{d_2}+\cdots+f_tx_n^{d_t}\in\Z[x_1,x_2,\dots,x_n]$,
$f_i \in \Z[x_1,x_2,\dots,x_{n-1}]$, $\deg(f)\leq D$, $C\geq \|f\|_\infty$, and $T\geq \#f$.
If $\beta_1\geq 2C+1,\beta_2\geq \beta_1^{D+1},\dots,\beta_n\geq \beta_{n-1}^{D+1}$, then
 $|f_i(\beta_1,\beta_2,\dots,\beta_{n-1})|\leq C\beta^D_{n-1}\frac{\beta_1-\frac{1}{\beta_1^{T-1}}}{\beta_1-1},i=1,2,\dots,t$.
\end{lemma}
\proof
It is easy to see that $T,D,C$ are also the corresponding bounds of $f_i$. Assume that $f_i=c_1m_1+c_2m_2+\dots+c_sm_s$, and $m_1<m_2<\cdots<m_s$ in lexicographic order. By Lemma \ref{lm-5}, we have $|f_i(\beta_1,\beta_2,\dots,\beta_{n-1})|\leq C(\widehat{m}_1+\widehat{m}_2+\cdots+\widehat{m}_s)=C\widehat{m}_s(\frac{\widehat{m}_1}{\widehat{m}_s}+\frac{\widehat{m}_2}{\widehat{m}_s}+\cdots+\frac{\widehat{m}_{s-1}}{\widehat{m}_s}+1)\leq C\beta^D_{n-1}(1+\frac{1}{\beta_1}+\frac{1}{\beta^2_1}+\cdots+\frac{1}{\beta_1^{T-1}})=C\beta^D_{n-1}\frac{\beta_1-\frac{1}{\beta_1^{T-1}}}{\beta_1-1}$.
\qed

Since $2|f_i(\beta_1,\beta_2,\dots,\beta_{n-1})|\leq 2C\beta^D_{n-1}\frac{\beta_1-\frac{1}{\beta_1^{T-1}}}{\beta_1-1}< \beta_{n-1}^D \beta_1\leq \beta_{n-1}^{D+1}\leq \beta_n$,
  $2|f_i(\beta_1,$ $\beta_2,\dots,$  $\beta_{n-1})|+1\leq \beta_n$, we can give the following recursive interpolation algorithm.
Note that we regard the upper bound $C\geq \|f\|_\infty$ to be a fixed number in the recursive process.

In order to be used in rational function interpolation algorithms, we denote the $f(\beta_1,\beta_2,\dots,\beta_n)$ as $\rho$ in the input of the following algorithm. In rational function interpolation, $\rho$ is $\frac{f(\beta_1,\beta_2,\dots,\beta_n)}{\mu}$ for some integer $\mu$. When $\mu=1$, Algorithm \ref{alg-11} always return the correct $f$.

\begin{alg}[MPolySIMod]\label{alg-11}
\end{alg}

{\noindent\bf Input:} A list $\beta_1,\beta_2,\dots,\beta_n$ in $\N$ which satisfies the condition of Lemma \ref{lm-202}; $\rho\in\Z$; $T,D\in\N$, where $T\geq \#f,D\geq \deg(f)$.

{\noindent\bf Output:} The exact form of $f(x_1,x_2,\dots,x_n)$ or failure.

\begin{description}
\item[Step 1:]
$\mathbf{If}$ $n=1$, $\mathbf{then}$ $C_1:=C$,  $\mathbf{else}$ $C_1:=\lfloor C\beta^{D}_{n-1}\frac{\beta_1-\frac{1}{\beta_1^{T-1}}}{\beta_1-1}\rfloor$, $\mathbf{end\ if}$;

\item[Step 2:]

Let $g:=\mathbf{UPolySIMod}(\beta_n,\rho,C_1,x_n)$;

$\mathbf{if}$ ($g=failure$ or $\deg(g)>D$) $\mathbf{then}$ return $failure$; $\mathbf{end\ if}$;

Assume $g=c_1x_n^{d_1}+c_2x_n^{d_2}+\cdots+c_tx_n^{d_t}$.

\item[Step 3:]
$\mathbf{If}$ $n=1$, $\mathbf{then}$ return $g$;

\item[Step 4:] Let $f:=0$;

\item[Step 5:]

for $i=1,2,\dots,t$ do

Let $M:=\mathbf{MPolySIMod}(\beta_1,\beta_2,\dots,\beta_{n-1},c_i,T-t+1,D-d_i)$.

$\mathbf{if}$  $M=$failure $\mathbf{then}$ $\mathbf{return}$ failure; $\mathbf{end\ if}$

$f:=f+Mx_n^{d_i}$;

\item[Step 6:]
$\mathbf{return}$ $f$.
\end{description}

\begin{theorem}\label{the-13}
The algorithm is correct.
The arithmetic complexity is $\mathcal{O}(nT\log^2 D)$, and the height of the data is $\mathcal{O}(D^n\log C)$.
\end{theorem}
\proof
By Theorem \ref{the-14}, the arithmetic operations of Algorithm $\mathbf{UPolySIMod}$ are $\mathcal{O}(T\log^2 D)$, and we call $n$ times Algorithm $\mathbf{UPolySIMod}$, so the arithmetic operations are $\mathcal{O}(nT\log^2 D)$.
The reason for the height of the data is the same as Theorem \ref{the-14}.\qed

\subsection{A probabilistic multivariate rational function interpolation algorithm}

In this and the next subsection, we assume
$$h=f/g\in \Z(x_1,x_2,\dots,x_n), \gcd(f,g)=1, T\geq \# h, D\geq \deg(h), C\geq \|h\|_\infty$$
and give a probabilistic algorithm.
We first prove a lemma.

\begin{lemma}
Assume $f,g\in \Z[x_1,x_2,\dots,x_n]$, $\gcd(f,g)=1$. If  $k_1,k_2,\dots,k_n$ are any positive numbers, then
$\gcd(f(x_1^{k_1},x_2^{k_2},\cdots,x_n^{k_n}),g(x_1^{k_1},x_2^{k_2},\cdots,x_n^{k_n}))=1$.
\end{lemma}
\proof
Since $\gcd(f,g)=1$, $\res(f,g,x_i)\neq 0$ for $i=1,2,\cdots,n$.
Denote $h_i:=\res(f,g,x_i)=s_if+t_ig$, where $h_i\in\Z[x_1,x_2,\dots,x_{i-1},x_{i+1},\dots,x_n]$.
Replacing $x_1,x_2,\dots,x_n$ by $x_1^{k_1},x_2^{k_2},\dots,$ $x_n^{k_n}$,
we have
$h_i(x_1^{k_1}, \dots,x_n^{k_n})=s_i(x_1^{k_1}, \dots,x_n^{k_n})f(x_1^{k_1},
\dots,x_n^{k_n})+t_i(x_1^{k_1}, \dots,x_n^{k_n})g(x_1^{k_1}, \dots,x_n^{k_n})$.
So $\gcd(f(x_1^{k_1},x_2^{k_2},\dots,x_n^{k_n}),g(x_1^{k_1},x_2^{k_2},\dots,x_n^{k_n}))|h_i(x_1^{k_1},x_2^{k_2},\dots,x_n^{k_n}),i=1,2,\dots,n$.
Since $h_i\neq 0$, it is easy to see that $h_i(x_1^{k_1},x_2^{k_2},\cdots,x_n^{k_n})\neq 0$, and we know $h_i(x_1^{k_1},x_2^{k_2},\cdots,x_n^{k_n})$ does not contain $x_i$.
So $\gcd(f(x_1^{k_1},x_2^{k_2},\cdots,x_n^{k_n}),g(x_1^{k_1},x_2^{k_2},\cdots,x_n^{k_n}))$ does not contain $x_1,x_2,$ $\cdots,x_n$. So we have $\gcd(f(x_1^{k_1},x_2^{k_2},\cdots,x_n^{k_n}),g(x_1^{k_1},x_2^{k_2},\cdots,x_n^{k_n}))=1$.\qed

\begin{lemma}
For $f,g\in \Z[x_1,x_2,\dots,x_n]$ and $\gcd(f,g)=1$,
 we have $\gcd(f(x_1+x,x_2+x,\dots,x_n+x),g(x_1+x,x_2+x,\dots,x_n+x))=1$
\end{lemma}
\proof
Let $h_i(x_1,x_2,\dots,x_{i-1},x_{i+1},\dots,x_n):=\res(f,g,x_i)=s_if+t_ig,i=1,2,\dots,n$.
From  $\gcd(f,g)=1$, we have $h_i\neq 0$.
Replacing $x_1,x_2,\dots,x_n$ by $x_1+x,x_2+x,\dots,x_n+x$,  we have $\gcd(f(x_1+x,x_2+x,\dots,x_n+x),g(x_1+x,x_2+x,\dots,x_n+x))|h_i(x_1+x,x_2+x,\dots,x_{i-1}+x,x_{i+1}+x,\dots,x_n+x)$.
Since $h_i(x_1+x,x_2+x,\dots,x_{i-1}+x,x_{i+1}+x,\dots,x_n+x)$ does not contain $x_i$,
$\gcd(f(x_1+x,x_2+x,\dots,x_n+x),g(x_1+x,x_2+x,\dots,x_n+x))$ contains variable $x$ only.

Denote $u(x):=\gcd(f(x_1+x,x_2+x,\dots,x_n+x),g(x_1+x,x_2+x,\dots,x_n+x))$. Then
$f(x_1+x,x_2+x,\dots,x_n+x)=u(x)a,g(x_1+x,x_2+x,\dots,x_n+x)=u(x)b$,
where $a,b\in \Z[x,x_1,x_2,\dots,x_n]$.
Regard $f(x_1+x, \dots,x_n+x),u(x),a(x,x_1, \dots,x_n),b(x,x_1, \dots,x_n)$ as polynomials in
$\C[x,x_1, \dots,x_n]$. If $u(x)$ is not a nonzero constant number, then let $\beta$ be a root of $u(x)$, and we have $f(x_1+\beta,x_2+\beta,\dots,x_n+\beta)=u(\beta)a(\beta,x_1,x_2,\dots,x_n)=0$.
Since the terms not containing variate $x$ in $f$ are the same as the the ones  in $f(x_1+\beta,x_2+\beta,\dots,x_n+\beta))$,   $f(x_1+\beta,x_2+\beta,\dots,x_n+\beta)\neq 0$.
This is a contradiction. So $u(x)$ is a nonzero number. So $\gcd(f(x_1+x,x_2+x,\dots,x_n+x),g(x_1+x,x_2+x,\dots,x_n+x))=1$.\qed

\begin{theorem}\label{lm-8}
Let $f,g\in \Z[x_1,x_2,\dots,x_n]$, $\gcd(f,g)=1$,
$D\geq\max\{\deg(f),\deg(g)\}$, $x,c_1,c_2,$ $\dots,c_n$ new variables. Then we have
$\gcd(f((x+c_1),(x+c_2)^{D+1},\dots,(x+c_n)^{(D+1)^{n-1}}),g((x+c_1),(x+c_2)^{D+1},\dots,(x+c_n)^{(D+1)^{n-1}})=1$.
\end{theorem}
\proof
By the two lemmas above, we can easyly obtain the theorem.\qed

By Theorem \ref{lm-8},  $R=\res(f(x+c_1,(x+c_2)^{D+1},\dots,(x+c_n)^{(D+1)^{n-1}}),g(x+c_1,(x+c_2)^{D+1},\dots,(x+c_n)^{(D+1)^{n-1}}),x)$ is a nonzero polynomial about $c_1,c_2,\dots,c_n$.
Then when we randomly choose $c_1,c_2,\dots,c_n$, with high probability, that $R(c_1,c_2,\dots,c_n)\ne0$.
So we reduce the multivariate case into univariate case. But the procedure will destroy the sparse structure.
In order to avoid this problem, we randomly choose $c_1,c_2,\dots,c_n$ satisfying $c_1\leq c_2\leq \cdots\leq c_n$, and then randomly choose a $\beta\geq 2C+1$ and
let $\beta_1=\beta+c_1,\beta_2=(\beta+c_2)^{D+1},\dots,\beta_n=(\beta+c_n)^{(D+1)^{n-1}}$.
Then these $\beta_i$ satisfy the condition of Theorem \ref{the-9} and the sparse structure is kept.

Assume $h(\beta_1,\beta_2,\dots,\beta_n)=\frac{a}{b}$, $\gcd(a,b)=1$, $a=\frac{f(\beta_1,\beta_2,\dots,\beta_n)}{\mu},b=\frac{g(\beta_1,\beta_2,\cdots,\beta_n)}{\mu}$.
We will give a bound of $|\mu|$.
\begin{lemma}
Let $h=\frac{f}{g}\in\Z(x_1,x_2,\dots,x_n)$,  $x_1^{e_1}x_2^{e_2}\cdots x_n^{e_n}$   the leading term of $f$, $c_1\leq c_2\leq\cdots\leq c_n$ positive integers. Let $\beta_1=2C+1+c_1,\beta_2=(2C+1+c_2)^{D+1},\dots,\beta_n=(2C+1+c_n)^{(D+1)^{n-1}}$ and $h(\beta_1,\beta_2,\cdots,\beta_n)=\frac{a}{b}$, then for every $j=1,2,\dots,n$, we have
$|\mu|< \frac{\beta_j\cdot \beta_j^{e_j}\beta_{j+1}^{e_{j+1}}\cdots \beta_n^{e_n}}{2|a|}$.
\end{lemma}

\proof
By Lemma \ref{the-9}, $\frac{|f(\beta_1,\beta_2,\cdots,\beta_n)|}{\beta_j^{e_j+1}\beta_{j+1}^{e_{j+1}}\cdots \beta_n^{e_n}}<\frac12$. Then we have
$\frac{\frac{1}{|\mu|} |f(\beta_1,\beta_2,\cdots,\beta_n)|}{\beta_j^{e_j+1}\beta_{j+1}^{e_{j+1}}\cdots \beta_n^{e_n}}<\frac1{2|\mu|}$, this is to say, $\frac{|a|}{\beta_j^{e_j+1}\beta_{j+1}^{e_{j+1}}\cdots \beta_n^{e_n}}<\frac1{2|\mu|}$, then we have $|\mu|< \frac{\beta_j^{e_j+1}\beta_{j+1}^{e_{j+1}}\cdots \beta_n^{e_n}}{2|a|}$.\qed

\begin{lemma}
$\frac{|\beta_j^{e_j+1}\beta_{j+1}^{e_{j+1}}\cdots \beta_n^{e_n}|}{2|a|}< |\mu| \beta_j$
\end{lemma}
\proof
$\frac{\beta_j^{e_j+1}\beta_{j+1}^{e_{j+1}}\cdots \beta_n^{e_n}}{2|a|}=\frac{\beta_j^{e_j+1}\beta_{j+1}^{e_{j+1}}\cdots \beta_n^{e_n}}{2\frac{1}{|\mu|}\cdot |f(\beta_1,\beta_2,\cdots,\beta_n)|}<|\mu| \beta_j\frac{\beta_j^{e_j}\beta_{j+1}^{e_{j+1}}\cdots \beta_n^{e_n}}{2|f(\beta_1,\beta_2,\cdots,\beta_n)|}< |\mu| \beta_j$\qed

By this lemma, if we know more about the degree of the leading terms,
and we can obtain smaller bounds of $|\mu|$.
\begin{lemma}\label{lm-9}
Suppose $h=\frac{f}{g}\in\Z(x_1,x_2,\dots,x_n)$.
Let $c_1\leq c_2\leq \dots\leq c_n$ be positive integers such that $\gcd(f(x+c_1,(x+c_2)^{2D+1},\dots,(x+c_n)^{(2D+1)^{n-1}}),g(x+c_1,(x+c_2)^{2D+1},\dots,(x+c_n)^{(2D+1)^{n-1}}))=1$, and $\beta\geq 2TC^2+1$. Then there is a unique $h(x_1,x_2,\dots,x_n)$ with $C\geq\|h\|_\infty$ corresponding to $h(\beta+c_1,(\beta+c_2)^{2D+1},\dots,(\beta+c_n)^{(2D+1)^{n-1}})$.
\end{lemma}
\proof
When $\gcd(f(x+c_1,(x+c_2)^{2D+1},\dots,(x+c_n)^{(2D+1)^{n-1}}),g(x+c_1,(x+c_2)^{2D+1},\dots,(x+c_n)^{{(2D+1)}^{n-1}}))=1$, $h(x_1,x_2,\dots,x_n)$ is in one-to-one correspondence with $h(x+c_1,(x+c_2)^{2D+1},\dots,(x+c_n)^{(2D+1)^{n-1}})$. If  $h(x+c_1,(x+c_2)^{2D+1},\dots,(x+c_n)^{(2D+1)^{n-1}})$ is unique, then   $h(x_1,x_2,\dots,x_n)$ is unique.

Assume there exists another rational function $\frac{f_1(x_1,x_2,\dots,x_n)}{g_1(x_1,x_2,\dots,x_n)}$ with $C\geq \|\frac{f_1}{g_1}\|_\infty$ such that
$\frac{f(\beta+c_1,(\beta+c_2)^{2D+1},\dots,(\beta+c_n)^{(2D+1)^{n-1}})}{g(\beta+c_1,(\beta+c_2)^{2D+1},\dots,(\beta+c_n)^{(2D+1)^{n-1}})}=\frac{f_1(\beta+c_1,(\beta+c_2)^{2D+1},\dots,(\beta+c_n)^{(2D+1)^{n-1}})}{g_1(\beta+c_1,(\beta+c_2)^{2D+1},\dots,(\beta+c_n)^{(2D+1)^{n-1}})}$,
which can be changed into $f(\beta+c_1,(\beta+c_2)^{2D+1},\dots,(\beta+c_n)^{(2D+1)^{n-1}})g_1(\beta+c_1,(\beta+c_2)^{2D+1},\dots,(\beta+c_n)^{(2D+1)^{n-1}})=g(\beta+c_1,(\beta+c_2)^{2D+1},\dots,(\beta+c_n)^{(2D+1)^{n-1}})f_1(\beta+c_1,(\beta+c_2)^{2D+1},\dots,(\beta+c_n)^{(2D+1)^{n-1}})$.

Define $w(x_1,x_2,\dots,x_n):=f(x_1,x_2,\dots,x_n)g_1(x_1,x_2,\dots,x_n)$ and $v(x_1,x_2,\dots,x_n):=f_1(x_1,x_2,\dots,x_n)g(x_1,x_2,\dots,x_n)$. Then $w(\beta+c_1,(\beta+c_2)^{2D+1},\dots,(\beta+c_n)^{(2D+1)^{n-1}})=v(\beta+c_1,(\beta+c_2)^{2D+1},\dots,(\beta+c_n)^{(2D+1)^{n-1}})$.
Since $2D\geq\deg(w),\deg(v)$, $TC^2 \geq \|w\|_\infty,\|v\|_\infty$,
by Lemma \ref{the-9}, we have $w(x_1,x_2,\dots,x_n)=v(x_1,x_2,\dots,x_n)$, so $\frac{f(x_1,x_2,\dots,x_n)}{g(x_1,x_2,\dots,x_n)}=\frac{f_1(x_1,x_2,\dots,x_n)}{g_1(x_1,x_2,\dots,x_n)}$, and the lemma is proved.\qed

Now we can give a probability algorithm.
\begin{alg}[MRFunSI1]\label{alg-mr1}
\end{alg}

{\noindent\bf Input:} A black box   $h=\frac{f}{g}\in\Z(x_1,x_2,\dots,x_n)$, $D,T,C,N$, where $D\geq \deg(h), T\geq \#h, C\geq \|h\|_\infty$, $N$ is a big positive integer.

{\noindent\bf Output:}
The exact form of $h(x_1,x_2,\dots,x_n)$ or a wrong rational function.

\begin{description}

\item[Step 1:]
Let $\beta:=2TC^2+1$. Randomly choose $c_1,c_2,\dots,c_n\in \{1,2,\dots,N\}$ such that $c_1\leq c_2\leq\cdots\leq c_n$.
Let $\beta_1:=\beta+c_1,\beta_2:=(\beta+c_2)^{2D+1},\cdots,\beta_n:=(\beta+c_n)^{(2D+1)^{n-1}}$.

\item[Step 2:]
Evaluate $h(\beta_1,\beta_2,\cdots,\beta_n)$ and assume $h(\beta_1,\beta_2,\cdots,\beta_n)=\frac{a}{b}$ with $\gcd(a,b,)=1$.

\item[Step 3:] $i=1$;

\item[Step 4:]

$f:=\mathbf{MPolySIMod}(\beta_1,\beta_2,\dots,\beta_n, a\cdot i,T,D,C)$;

$\mathbf{if}$ $f=failure$ $\mathbf{then}$ $i:=i+1$; go to Step $4$; $\mathbf{end\ if}$

\item[Step 5:]

$g:=\mathbf{MPolySIMod}(\beta_1,,\beta_2,\dots,\beta_n,b\cdot i,T,D,C)$;

$\mathbf{if}$ $g=failure$ $\mathbf{then}$ $i:=i+1$; go to Step $4$; $\mathbf{end\ if}$

\item[Step 6:]
$\mathbf{Return}$ $\frac{f}{g}$.

\end{description}

\begin{theorem}\label{the-mr2}
The algorithm is correct.
The arithmetic complexity is $\mathcal{O}(\mu nT\log^2 D)$, and the height of the data is $\mathcal{O}((2D)^n\log (TC^2+N)$.
\end{theorem}
\proof
By Lemma \ref{lm-9}, if $\gcd(f(x+c_1,(x+c_2)^{2D+1},\dots,(x+c_n)^{(2D+1)^{n-1}}),g(x+c_1,(x+c_2)^{2D+1},\dots,(x+c_n)^{(2D+1)^{n-1}}))=1$, then we can find a rational function with coefficients bounded by $C$ only when $i=\mu$. So in this case, the algorithm returns a correct $h$.
Otherwise, it may return a wrong rational function.

By Theorem \ref{the-14}, the arithmetic complexity of Algorithm $\mathbf{UPolySIMod}$ is $\mathcal{O}(T\log^2 D)$ and we call $n$ times Algorithm $\mathbf{UPolySIMod}$, so the arithmetic complexity of $\mathbf{MPolySIMod}$ is $\mathcal{O}(nT\log^2 D)$. 
The algorithm calls algorithm $\mathbf{MPolySIMod}$ at most $\mu$ times, so the arithmetic complexity is $\mathcal{O}(\mu nT\log^2 D)$.
The reason for the height of the data is the same as Lemma \ref{the-14}. In this case, the degree is $O((2D)^n)$, and $\beta$ is $O(TC^2+N)$. So the height of the data is $\mathcal{O}(D^n\log (TC^2+N)$.\qed

We now analyze the successful rate  of Algorithm \ref{alg-mr1}.

\begin{lemma}\label{lm-14}
Let $R$ be an integral domain, $S_1,S_2,\dots,S_n\subseteq R$ finite sets with $N=\#S_i,i=1,2,\dots,n$ elements, and $r\in R[x_1,x_2,\dots,x_n]$ a polynomial of total degree at most $d\in \N$. If $r$ is not the zero polynomial, then $r$ has at most $dN^{n-1}$ zeros in $S_1\times S_2\times \cdots\times S_n$.
\end{lemma}
\proof
 We prove it by induction on $n$. The case $n=1$ is clear, since a nonzero univariate polynomial of degree at most $d$ over an integral has at most $d$ zeros. For the induction step, we write $r$ as a polynomial in $x_n$: $r=\sum_{0\leq i\leq k}r_ix_n^i$ with $r_i\in R[x_1,x_2,\dots,x_{n-1}]$ for $0\leq i\leq k$ and $r_k\neq 0$. Then $\deg(r_k)\leq d-k$. By the induction hypothesis, $r_k$ has at most $(d-k)N^{n-2}$ zeroes in $S_1\times S_2\times \cdots\times S_{n-1}$. So that there are at most $(d-k)N^{n-1}$ common zeroes of $r$ and $r_k$ in $S_1\times S_2\times \cdots\times S_n$. Furthermore, for each $a\in S_1\times S_2\times \cdots S_{n-1}$ with $r_k(a)\neq 0$, the univariate polynomial $r_a=\sum_{0\leq i\leq k}r_i(a)x_n^i\in R[x_n]$ of degree $k$ has at most $k$ zeros, so that the total number of zeros of $r$ in $S^n$ is bound by $(d-k)N^{n-1}+kN^{n-1}=dN^{n-1}$ .\qed

\begin{theorem}
$S_1,S_2,\dots,S_n$ are $n$ different positive integer sets with $\#S_i=N$. Assume $a_i<a_j$ when $i<j$ where $a_i$ is any elements in $S_i$. If $c_1,c_2,\dots,c_n$ are randomly chosen in $S_1\times S_2\cdots\times S_n$, then Algorithm \ref{alg-mr1} returns the correct result with probability at least $1-\frac{2(2D+1)^{2n}}{N}$.
\end{theorem}
\proof
Denote $f_0 = f(x+c_1,(x+c_2)^{2D+1},\dots,(x+c_n)^{(2D+1)^{n-1}})$,
$g_0= g(x+c_1,(x+c_2)^{2D+1},\dots,(x+c_n)^{(2D+1)^{n-1}})$.
By Lemma \ref{lm-9},  when $\gcd(f_0,g_0)=1$ in the above algorithm, we obtain the correct result.
By Lemma \ref{lm-8}, we know $\res(f_0,g_0,x)\neq0$.
We can see $\deg_x f_0<(2D+1)^n,\deg_x g_0<(2D+1)^n$,
$\deg_{c_1,c_2,\dots,c_n} f_0<(2D+1)^n, \deg_{c_1,c_2,\dots,c_n} g_0<(2D+1)^n$, so $\deg_{c_1,c_2,\dots,c_n}\res(f_0,g_0,x)<2(2D+1)^{2n}$.
By Lemma \ref{lm-14}, if $c_1,c_2,\dots,c_n$ are randomly chosen from $S_1\times S_2\cdots \times S_n$, then the probability of resultant polynomial be zero at point $(c_1,c_2,\dots,c_n)$ is no more than $\frac{2(2D+1)^{2n}}{N}$.
So the success rate of Algorithm \ref{alg-mr1} is at least $1-\frac{2(2D+1)^{2n}}{N}$.\qed

\subsection{A probabilistic algorithm with two smaller sample points}
In this section, we give a new algorithm based on two evaluations,
which is a combination of Algorithms \ref{alg-mr1} and \ref{ProbUniRFunInt}.

\begin{lemma}\label{lm-6}
Let $f=c_1m_1+c_2m_2+\cdots+c_tm_t\in Z[x_1,x_2,\dots,x_n],m_1<m_2<\cdots<m_t$, $m_t=x_1^{e_1}x_2^{e_2}\cdots x_n^{e_n}$, $D\geq \deg(f),C\geq \|f\|_\infty$.
Let $\beta\geq 2C+1,\beta_1\geq \beta,\beta_2\geq \beta_1^{D+1},\cdots,\beta_n\geq \beta_{n-1}^{D+1}$,  $Q:=\frac{f(\beta_1,\beta_2,\cdots,\beta_n)/\beta_n^{e_n}}{f(\beta_1,\beta_2,\cdots,\beta_n+1)/(\beta_n+1)^{e_n}},E:=1+\frac{2C}{(\beta_1-1)(\beta_n-1)}$, then we have
$\frac{1}{E}<Q<E$.
\end{lemma}

\proof
Denote $\widehat{m}^1_i,\widehat{m}^2_i$ to be the values of $m_i$ by replacing $x_1,x_2,\cdots,x_n$  with $\beta_1,\beta_2,\cdots,\beta_n$ and $\beta_1,\beta_2,\cdots,\beta_n+1$, ,$i=1,2,\cdots,t$, respectively.
Let $q_1:=\frac{c_1\widehat{m}^1_1+c_2\widehat{m}^1_2+\cdots+c_{t-1}\widehat{m}^1_{t-1}}{\widehat{m}^1_t},q_2:=\frac{c_1\widehat{m}^2_1+c_2\widehat{m}^2_2+\cdots+c_{t-1}\widehat{m}^2_{t-1}}{\widehat{m}^2_t}$. Then $Q=\frac{f(\beta_1,\beta_2,\cdots,\beta_n)/\widehat{m}^1_t}{f(\beta_1,\beta_2,\cdots,\beta_n+1)/\widehat{m}^2_t}=\frac{c_t+q_1}{c_t+q_2}=1+\frac{q_1-q_2}{c_t+q_2}$.
Since $|c_t|\geq 1,|q_2|<\frac{\varepsilon}{2}=\frac{1}{2}$, we have $|c_t+q_2|> \frac{1}{2}$, and $|\frac{q_1-q_2}{c_t+q_2}|<2|q_1-q_2|$.
$|q_1-q_2|=|\frac{c_1\widehat{m}^1_1+c_2\widehat{m}^1_2+\cdots+c_{t-1}\widehat{m}^1_{t-1}}{\widehat{m}^1_t}-\frac{c_1\widehat{m}^2_1+c_2\widehat{m}^2_2+\cdots+c_{t-1}\widehat{m}^2_{t-1}}{\widehat{m}^2_t}|$
$\leq C|\frac{\widehat{m}_1^1}{\widehat{m}_t^1}-\frac{\widehat{m}_1^2}{\widehat{m}_t^2}|+|\frac{\widehat{m}_2^1}{\widehat{m}_t^1}-\frac{\widehat{m}_2^2}{\widehat{m}_t^2}|+\cdots+C|\frac{\widehat{m}_{t-1}^1}{\widehat{m}_t^1}-\frac{\widehat{m}_{t-1}^2}{\widehat{m}_t^2}|$.

Consider the last summand $|\frac{\widehat{m}_{t-1}^1}{\widehat{m}_t^1}-\frac{\widehat{m}_{t-1}^2}{\widehat{m}_t^2}|$.
We assume $m_{t-1}=x_1^{t_1}x_2^{t_2}\cdots x_n^{t_n}$. Then
$|\frac{\widehat{m}_{t-1}^1}{\widehat{m}_t^1}-\frac{\widehat{m}_{t-1}^2}{\widehat{m}_t^2}|=|\frac{\beta_1^{t_1}\beta_2^{t_2}\cdots \beta_n^{t_n}}{\beta_1^{e_1}\beta_2^{e_2}\cdots \beta_n^{e_n}}-\frac{\beta_1^{t_1}\beta_2^{t_2}\cdots (\beta_n+1)^{t_n}}{\beta_1^{e_1}\beta_2^{e_2}\cdots (\beta_n+1)^{e_n}}|=\frac{\beta_1^{t_1}\beta_2^{t_2}\cdots \beta_{n-1}^{t_{n-1}}}{\beta_1^{e_1}\beta_2^{e_2}\cdots \beta_{n-1}^{e_{n-1}}}|\frac{1}{\beta_n^{e_n-t_n}}-\frac{1}{(\beta_n+1)^{e_n-t_n}}|$.
Since $m_{t-1}<m_t$, we have $t_n\leq e_n$.  If $t_n=e_n$, then the summand will be zero, so we need only consider the case $t_n<e_n$.
It is easy to see that $\beta_1^{t_1}\beta_2^{t_2}\cdots \beta_{n-1}^{t_{n-1}}\leq \frac{\beta_n}{\beta_1}$. So we have
{\tiny
\begin{align}
\frac{1}{C}|q_1-q_2|\leq{} & \frac{\beta_n}{\beta_1^{e_1}\beta_2^{e_2}\cdots \beta_{n-1}^{e_{n-1}}}(\frac{1}{\beta_1}+\frac{1}{\beta_1^2}\cdots+\frac{1}{\beta_1^T})(\frac{1}{\beta_n}-\frac{1}{\beta_n+1})\notag\\
+{}&\frac{\beta_n}{\beta_1^{e_1}\beta_2^{e_2}\cdots \beta_{n-1}^{e_{n-1}}}(\frac{1}{\beta_1}+\frac{1}{\beta_1^2}+\cdots+\frac{1}{\beta_1^T})(\frac{1}{\beta_n^2}-\frac{1}{(\beta_n+1)^2})\notag\\
{}&                   \vdots\notag\\
+{}&\frac{\beta_n}{\beta_1^{e_1}\beta_2^{e_2}\cdots \beta_{n-1}^{e_{n-1}}}(\frac{1}{\beta_1}+\frac{1}{\beta_1^2}+\cdots+\frac{1}{\beta_1^T})(\frac{1}{\beta_n^D}-\frac{1}{(\beta_n+1)^D})\notag\\
={}&\frac{\beta_n}{\beta_1^{e_1}\beta_2^{e_2}\cdots \beta_{n-1}^{e_{n-1}}} \frac{1-\frac{1}{\beta_1^T}}{\beta_1-1}((\frac{1}{\beta_n}+\frac{1}{\beta_n^2}+\cdots+\frac{1}{\beta_n^D})-(\frac{1}{\beta_n+1}+\frac{1}{(\beta_n+1)^2}+\cdots+\frac{1}{(\beta_n+1)^D}))\notag\\
={}&\frac{\beta_n}{\beta_1^{e_1}\beta_2^{e_2}\cdots \beta_{n-1}^{e_{n-1}}} \frac{1-\frac{1}{\beta_1^T}}{\beta_1-1}(\frac{1-\frac{1}{\beta_n^D}}{\beta_n-1}-\frac{1-\frac{1}{(\beta_n+1)^D}}{\beta_n})\notag\\
\leq{} & \frac{1}{\beta_1^{e_1}\beta_2^{e_2}\cdots \beta_{n-1}^{e_{n-1}}}\frac{1}{\beta_1-1}\frac{1}{\beta_n-1}\leq \frac{1}{\beta_1-1}\frac{1}{\beta_n-1}.\notag
\end{align}}
So $|q_1-q_2|\leq C\frac{1}{\beta_1-1}\frac{1}{\beta_n-1}$ and hence
\begin{equation}
\begin{cases}
1-\frac{2C}{(\beta_1-1)(\beta_n-1)}<Q< 1+\frac{2C}{(\beta_1-1)(\beta_n-1)}\\
1-\frac{2C}{(\beta_1-1)(\beta_n-1)}<\frac{1}{Q}< 1+\frac{2C}{(\beta_1-1)(\beta_n-1)}
\end{cases}
\end{equation}
\qed

\begin{lemma}\label{lm-7}
Let $h=\frac{f}{g}\in \Z(x_1,x_2,\dots,x_n)$, $D\geq \deg(f)$, $D_n\geq \deg_{x_n}(f)\geq d_n$. Let $\beta\geq 2C+1,\beta_1\geq \beta,\beta_2\geq \beta_1^{D+1},\dots,\beta_n\geq \beta_{n-1}^{D+1}$, $h(\beta_1,\beta_2,\dots,\beta_n)=\frac{a_1}{b_1},h(\beta_1,\beta_2,\dots,\beta_n+1)=\frac{a_2}{b_2}$, $a_1=\frac{f(\beta_1,\beta_2,\dots,\beta_n)}{\mu_1},a_2=\frac{f(\beta_1,\beta_2,\dots,\beta_n+1)}{\mu_2}$, and $Q_1:=\frac{a_1/\beta_n^{d_n}}{a_2/(\beta_n+1)^{d_n}},Q_2:=\frac{a_1/\beta_n^{D_n}}{a_2/(\beta_n+1)^{D_n}},E:=1+\frac{2C}{(\beta_1-1)(\beta_n-1)}$.
Then
$|Q_1|\frac{1}{E}< |\frac{\mu_2}{\mu_1}|< |Q_2|E$.
\end{lemma}
\proof
This lemma can be proved similar to  Lemma \ref{lm-3}.\qed

Now we give the algorithm which is similarly to Algorithm \ref{ProbUniRFunInt}.
\begin{alg}[MRFunSI2]
\end{alg}

{\noindent\bf Input:} A black box $h=\frac{f}{g}\in\Z(x_1,x_2,\dots,x_n)$, $D,D_n,C,N$, where $D\geq \deg(h),D_n\geq \deg_{x_n}(f),C\geq \|h\|_\infty$, $N$ a big positive integer.

{\noindent\bf Output:}  The exact form of $h(x_1,x_2,\dots,x_n)$ or a wrong rational function.

\begin{description}

\item[Step 1:]
Let $\beta:=3C+1$. Randomly choose $c_1,c_2,\dots,c_n\in \{1,2,\dots,N\}$ such that $c_1\leq c_2\leq \cdots\leq c_n$.
Let $\beta_1:=\beta+c_1,\beta_2:=(\beta+c_2)^{D+1},\cdots,\beta_n:=(\beta+c_n)^{(D+1)^{n-1}}$.

\item[Step 2:]
Evaluate $h(\beta_1,\beta_2,\cdots,\beta_n),h(\beta_1,\beta_2,\cdots,\beta_n+1)$, and assume $h(\beta_1,\beta_2,\cdots,\beta_n)=\frac{a_1}{b_1},h(\beta_1,\beta_2,\cdots,\beta_n+1)=\frac{a_2}{b_2}$.

\item[Step 3:]
%
Let $d:=\max(\lfloor \log_{\beta_n} (2a_1)\rfloor, \lfloor \log_{\beta_n+1} (2a_2)\rfloor$ (due to Lemma \ref{lm-1}).

\item[Step 4:]
Let $k_1:=\lfloor\frac{\beta_n^{D_n+1}}{|a_1|}\rfloor,k_2:=\lfloor\frac{(\beta_n+1)^{D_n+1}}{|a_2|}\rfloor$;
$Q_1:=|\frac{a_1/\beta_n^{d_n}}{a_2/(\beta_n+1)^{d_n}}|$
$Q_2:=|\frac{a_1/\beta_n^{D_n}}{a_2/(\beta_n+1)^{D_n}}|$
$E:=1+\frac{2C}{(\beta_1-1)(\beta_n-1)}$

\item[Step 5:]
Let $i:=1$.

$\mathbf{If}$ $Q_1\geq E$ $\mathbf{then}$ goto step 6;
$\mathbf{If}$ $Q_2\leq \frac{1}{E}$ $\mathbf{then}$ goto step 7;

$\mathbf{If}$ $k_1<k_2$, $\mathbf{then}$ goto step 6;
$\mathbf{Else}$
goto step 7.


\item[Step 6:]

$\mathbf{while}$ $i\leq k_1$ $\mathbf{do}$

$\mathbf{if}$ the interval $(\frac{Q_1}{E}i, Q_2Ei )$ includes an integer

 $\mathbf{then}$ $f:=\mathbf{MPolySIMod}(\beta_1,\beta_2,\dots,\beta_n, D, a_1\cdot i,C)$

 $\mathbf{if}$ $f = \mathbf{failure}$ $\mathbf{then}$ $i:=i+1$; goto step 6; $\mathbf{end\ if}$;

 $g:=\mathbf{MPolySIMod}(\beta_1,\beta_2,\dots,\beta_n, D, b_1\cdot i,C)$

 $\mathbf{if}$ $g = \mathbf{failure}$ $\mathbf{then}$ $i:=i+1$; goto step 6; $\mathbf{end\ if}$;

 $\mathbf{if}$ $h(\beta_1,\beta_2,\dots,(\beta_n+1))=\frac{f(\beta_1,\beta_2,\dots,(\beta_n+1))}{g(\beta_1,\beta_2,\dots,(\beta_n+1))}$ $\mathbf{then}$ return $\frac{f(x_1,x_2,\dots,x_n)}{g(x_1,x_2,\dots,x_n)}$.

\item[Step 7:]

$\mathbf{while}$ $i\leq k_2$ $\mathbf{do}$

$\mathbf{if}$ the interval $(\frac{1}{Q_2E}i, \frac{E}{Q_1}i )$ includes an integer

 $\mathbf{then}$ $f:=\mathbf{MPolySIMod}(\beta_1,\beta_2,\dots,\beta_n+1, D, a_2\cdot i,C)$

  $\mathbf{if}$ $f = \mathbf{failure}$ $\mathbf{then}$ $i:=i+1$; goto step 7; $\mathbf{end\ if}$;

 $g:=\mathbf{MPolySIMod}(\beta_1,\beta_2,\dots,\beta_n+1, D, b_2\cdot i,C)$

  $\mathbf{if}$ $f = \mathbf{failure}$ $\mathbf{then}$ $i:=i+1$; goto step 7; $\mathbf{end\ if}$;

 $\mathbf{if}$ $h(\beta_1,\beta_2,\dots,\beta_n)=\frac{f(\beta_1,\beta_2,\dots,\beta_n)}{g(\beta_1,\beta_2,\dots,\beta_n)}$  $\mathbf{then}$ return $\frac{f(x_1,x_2,\dots,x_n)}{g(x_1,x_2,\dots,x_n)}$.

\end{description}

\begin{remark}
 In the above algorithm, we also call algorithm $\mathbf{\mathbf{MPolySIMod}}$.
But the size of $\beta$ is reduced by a factor of $T$. So in Algorithm $\mathbf{MPolySIMod}$, we need adjust its step 1 as

 $\mathbf{If}$ $n=1$, $\mathbf{then}$ $C_1:=C$,  $\mathbf{else}$ $C_1:=\lfloor C\beta^{D}_{n-1}\frac{\beta_1}{\beta_1-1}\rfloor$, $\mathbf{end\ if}$;
\end{remark}

\section{Experimental results}
\label{sec-exp}

In this section, practical performances of the new algorithms will be presented.
The data are collected on a desktop with Windows system,
3.60GHz Core $i7-4790$ CPU, and 8GB RAM memory.
%


Five randomly constructed rational functions are used to obtian the average times.
We have four groups of experiments to present. The first and second groups are about univariate rational function interpolation. The third and fourth groups are about multivariate rational function interpolation.

\begin{figure}[!hptb]
\begin{minipage}[t]{0.49\linewidth}
\centering
\includegraphics[scale=0.25]{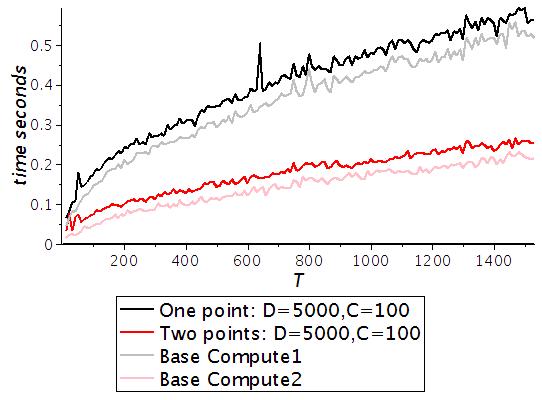}
\caption{Univariate:
average running times with varying $T$} \label{fig11}
\end{minipage}\quad
\begin{minipage}[t]{0.48\linewidth}
\centering
\includegraphics[scale=0.24]{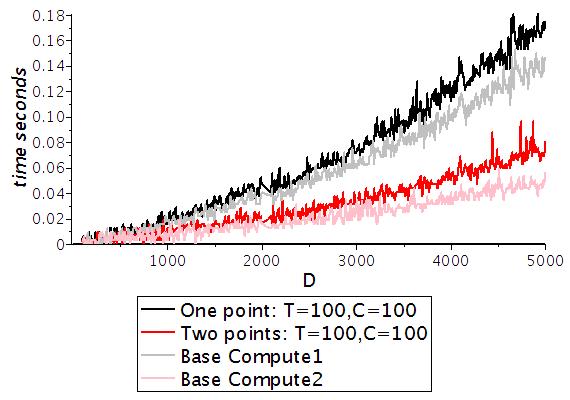}
\caption{Univariate:
average running times with varying $D$ }\label{fig12}
\end{minipage}
\end{figure}

In Figures \ref{fig11} and \ref{fig12},  we compare the two deterministic algorithms $\mathbf{URFunSI1}$ and $\mathbf{URFunSI2}$ for univariate rational function interpolation.
By the {\bf Base Case}, we mean the sum of the times of interpolating $f$ and $g$ separately.
From the data, we can see that
(1) the algorithm using two points are faster than that using one point and
(2) the times for interpolating $h=\frac{f}{g}$
are almost the same as that of interpolating $f$ and $g$, which means
that our interpolation algorithm for univariate rational functions are almost optimal.

\begin{figure}[!hptb]
\begin{minipage}[t]{0.49\linewidth}
\centering
\includegraphics[scale=0.22]{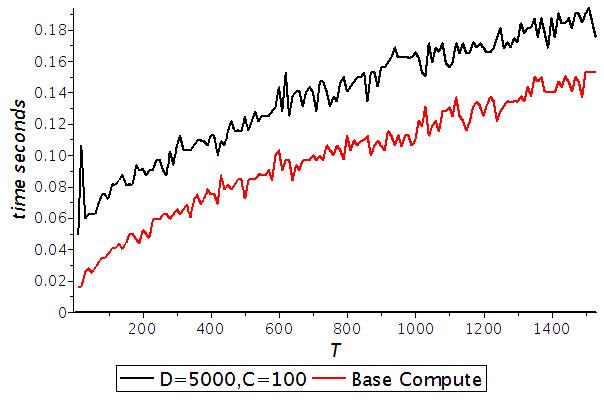}
\caption{$\mathbf{URFunSIP}$:
average running times with varying $T$} \label{fig13}
\end{minipage}\quad
\begin{minipage}[t]{0.48\linewidth}
\centering
\includegraphics[scale=0.21]{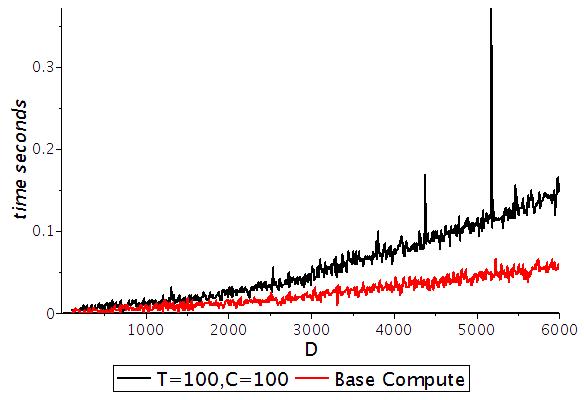}
\caption{$\mathbf{URFunSIP}$:
average running times with varying $D$ }\label{fig14}
\end{minipage}
\end{figure}

In Figures \ref{fig13} and \ref{fig14}, we present the practical performance
for the probabilistic algorithm for univariate rational functions. We compare it with the base case.
Comparing Figure \ref{fig11} and Figure \ref{fig13}, we can see that the probabilistic algorithm is faster than the one point deterministic algorithm and comparable with the
two points deterministic algorithm.

\begin{figure}[!hptb]
\begin{minipage}[t]{0.49\linewidth}
\centering
\includegraphics[scale=0.22]{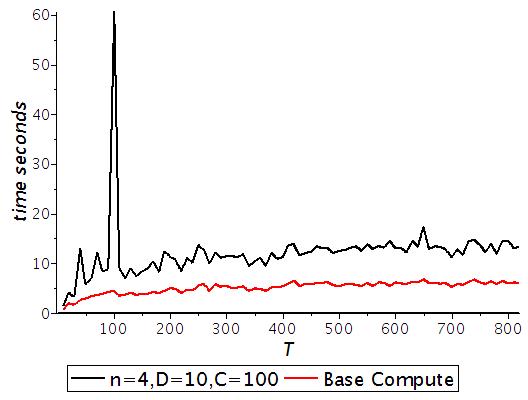}
\caption{$\mathbf{MRFunSIP1}$:
average running times with varying $T$} \label{fig15}
\end{minipage}\quad
\begin{minipage}[t]{0.48\linewidth}
\centering
\includegraphics[scale=0.21]{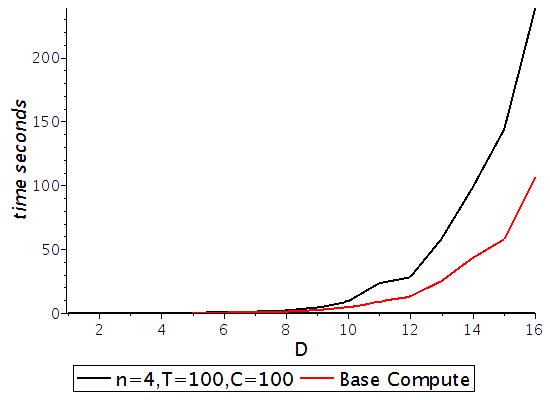}
\caption{$\mathbf{MRFunSIP1}$:
average running times with varying $D$ }\label{fig16}
\end{minipage}
\end{figure}

\begin{figure}[!hptb]
\begin{minipage}[t]{0.49\linewidth}
\centering
\includegraphics[scale=0.22]{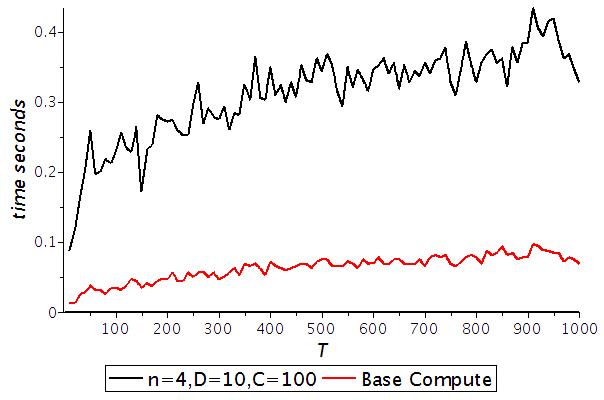}
\caption{$\mathbf{MRFunSIP2}$:
average running times with varying $T$} \label{fig17}
\end{minipage}\quad
\begin{minipage}[t]{0.48\linewidth}
\centering
\includegraphics[scale=0.21]{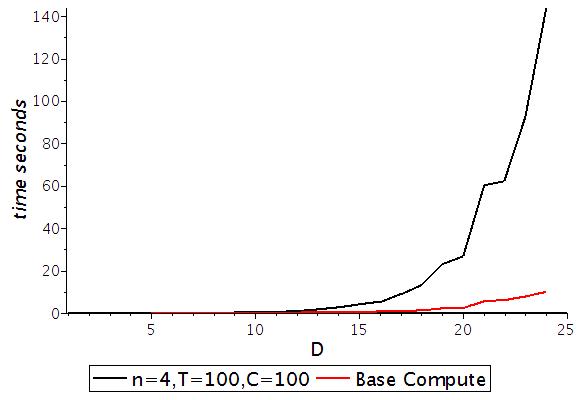}
\caption{$\mathbf{MRFunSIP2}$:
average running times with varying $D$ }\label{fig18}
\end{minipage}
\end{figure}

For the multivariate algorithm, in Figures \ref{fig15}, \ref{fig16}, \ref{fig17}, and \ref{fig18}, we present the practical performances with one point and with two points. We also give the time which is the sum of the times of interpolating $f$ and $g$  from $h=\frac{f}{g}$ for comparison.
We can see that the algorithm with two points is much better than the algorithm with one point.
Both algorithms are less sensitive to $T$ and are quite sensitive to $D$.
But unlike the univariate case, the interpolation of the rational function
is much difficult than interpolating its denominator and numerator separatively.

\section{Conclusion}
In this paper, we consider interpolation of sparse rational functions under the assumption that their coefficients are integers with a given bound.
This assumption allows us to recover the rational function $h=f/g$ from
evaluations of $h$ at one ``large'' sample point.
Experimental results show that the univariate interpolation algorithm is almost optimal,
while the multivariate interpolation algorithm needs further improvements.
The main problem is that the sample data is of exponential size in $n$.
The main reason is using of Kronecker type substitution.

\end{document}